\documentclass[]{myElsart}
\newcommand\independent{\protect\mathpalette{\protect\independenT}{\perp}}
\def\independenT#1#2{\mathrel{\rlap{$#1#2$}\mkern5mu{#1#2}}}

\newcommand{\E}{\mathbb{E}}
\newcommand{\Prob}{\mathbb{P}}

\usepackage[all]{xy}
\xyoption{arc}
\usepackage{amsmath}
\usepackage{amssymb}
\usepackage{bbm}
\usepackage{booktabs}
\usepackage{theorem}
\usepackage[]{natbib}
\usepackage{tikz}
\usetikzlibrary{arrows}
\usepackage{enumitem}

\newcounter{count}
\newenvironment{theorem}[1][Theorem \arabic{count}]{\vspace{1em}\refstepcounter{count}\begin{trivlist}
\item[\hskip \labelsep {\bfseries #1}]}{\end{trivlist}\vspace{1em}}

\newenvironment{proposition}[1][Proposition \arabic{count}]{\vspace{1em}\refstepcounter{count}\begin{trivlist}
\item[\hskip \labelsep {\bfseries #1}]}{\end{trivlist}\vspace{1em}}

\newenvironment{lemma}[1][Lemma \arabic{count}]{\vspace{1em}\refstepcounter{count}\begin{trivlist}
\item[\hskip \labelsep {\bfseries #1}]}{\end{trivlist}\vspace{1em}}

\newenvironment{corollary}[1][Corollary \arabic{count}]{\vspace{1em}\refstepcounter{count}\begin{trivlist}
\item[\hskip \labelsep {\bfseries #1}]}{\end{trivlist}\vspace{1em}}

\newenvironment{example}[1][Example \arabic{count}]{\vspace{1em}\refstepcounter{count}\begin{trivlist}
\item[\hskip \labelsep {\bfseries #1}]}{\end{trivlist}\vspace{1em}}

\newenvironment{proof}[1][Proof]{\vspace{1em}\begin{trivlist}
\item[\hskip \labelsep {\bfseries #1}]}{\hfill$\Box$\end{trivlist}\vspace{1em}}

\begin{document}

%---------------------------------------------------
\begin{frontmatter}

\title{Measurement error and reliability from a different perspective: Reliability of decision functions and the sum score}
%-------------------------------------------------------
\author{Lourens Waldorp,}
%\ead{waldorp@uva.nl}
\author{Maarten Marsman, and}
\author{Denny Borsboom}
\runauthor{Waldorp, Marsman and Borsboom}
\runtitle{Fourier analysis of decision functions of tests}
\address{University of Amsterdam, Nieuwe Achtergracht 129-B, 1018 NP, the Netherlands\\
{\tt waldorp@uva.nl \hspace{3em} m.marsman@uva.nl \hspace{3em} d.borsboom@uva.nl}}
%-------------------------------------------------------
\begin{abstract}
We propose an alternative framework for measurement error that allows us to determine reliability at the level of the decision, such as the decision to pass or fail. This framework makes it possible to answer the question which properties of a decision function are relevant to make decisions. The mechanism is relatively simple for dichotomous items: The response to any item has probability $\tfrac{1}{2}(1-\rho)$, with $\rho\in [-1,1]$, to be different from the original response. We show that this framework satisfies the classical axioms (those of Lord and Novick) and can therefore be considered as being aligned with classical test theory. We also show some relations to modern test theory and its connections to graphs. We then connect the idea of reliability at the level of the decision to the idea of how many of the (two-response) items are required to be flipped to change the decision; this is referred to as stability. Finally, we argue that from this perspective the best decision function (with respect to a set of properties including stability) is a weighted sum score with threshold value. 
\end{abstract}
%-------------------------------------------------------

\end{frontmatter}
%-------------------------------------------------------
\endNoHyper

%---------------------------------------------------------------------------
\section{Introduction}\noindent
Measurement error is traditionally defined in terms of a true score, to which some random variable is added such that, in expectation, the observed score equals the true score \citep[][Definition 2.4.1]{Lord:1968}; the difference between the observed score and the true score is then conceptualized as measurement error. This tradition has borne fruit to many theoretical results \citep[e.g.,][]{Holland:1986b,Junker:1993,Ellis:1997,Bork:2022,Sijtsma:2024} and many practical guidelines to create and evaluate tests \citep[e.g.,][]{Cronbach:1963,Sijtsma:1987}.However, it has remained somewhat poorly illuminated how items with measurement error should be aggregated (i.e., a sum score or harmonic mean) to obtain the best possible decision. 

Here we propose an alternative perspective on measurement error, which allows us to provide a systematic answer to the question of how to aggregate the items to reach a decision (with respect to properties deemed desirable by the test user). We define measurement error in terms of a random experiment on the original response such that, in the dichotomous case, the response is either flipped with probability $\tfrac{1}{2}(1-\rho)$ or is not flipped, where the parameter $\rho$ is in the interval $[-1,1]$. We show how this framework leads to the traditional axioms of classical test theory \citep[][Section 3.1]{Lord:1968}, and we are therefore able to determine the reliability of items. As a corollary of our framework, in which we use graphical models, viewing the items in a test as nodes in a graph that are connected in a specific way, gives, in our view, a useful perspective on how items relate to each other, and we will also show how they affect each other's reliability (Section \ref{sec:measurement-error-classical}). 
We then connect reliability to the sensitivity of a decision function (a rule based on the items to reach a decision) to measurement error, which tells us what fraction of items can be flipped such that the decision remains the same (stability). We claim that these concepts are important to determine \textit{how} decisions based on items are made.

Previous research with respect to such decision functions based on test items, is mostly about scoring rules; which weights should be assigned to items. Such research resulted in Chernoff's rule \citep[see e.g., ][Chap. 14]{Lord:1968}, optimal cut-off scores for a sum score to minimise Bayes' risk  \citep{Linden:1980,Linden:1987}, cut-off values on the scale of a latent variable such that sensitivity and specificity of the classification are as high as possible with respect to clinical diagnoses \citep{Goldberg:1988}, and, more recently, joint rules for accuracy and response time \citep{Maris:2012}. Here we are interested in other types of motivations for using a particular decision function. For instance, a most relevant property of a decision function, we believe, is stability: If we change a small fraction of the original data, then the decision should be unaffected. Stability of the decision will be shown to be directly related to reliability of the decision, a concept used for item analysis \citep{Linden:2018}. 

As a consequence of our proposed framework and our claim that stability is an important requirement for decision functions, we try to determine which kind of decision functions are best, with respect to a set of desiderata (see Section \ref{sec:merit-sum-score}). Traditionally, psychometric properties have been associated with the fit of a particular type of model, one that often relies on measuring constructs, which are defined in terms of latent variables \citep{Lord:1968,Holland:1986b,Junker:1993}. Such models, like  Item Response Theory (IRT) models, focus on inference with respect to a latent variable. The normative function of such models determines, for instance, whether an item is considered good or bad. If the model describes the test well, it can be used to make decisions, to pass or fail, for example. In a latent variable model, like the Rasch model, the sum score is a sufficient statistic \citep[][Chapter 3]{Rasch:1960,Linden:2013}; and as such the sum score is appropriate for inference with respect to the latent variable \citep[][Chapter 6]{Linden:2018}. A slightly different argument in favour of the sum score was recently put forward in terms of networks: the sum score appears to be strongly related to the expectation of all items \citep{Sijtsma:2024}. Although this result hinges on the fact that there is no bimodality in the latent variable distribution, this argument is more in line with what we try to achieve here. 

Our contributions to the vast field of test theory are a) introducing a new mechanism for measurement error, b) determining a set of relevant properties for a decision function to have, and c) extending Fourier analysis of Boolean functions to dependent variables (using graphical models) to analyse decision functions in realistic situations.

\subsection{Overview}\noindent
Our main objective is to determine a way to evaluate a decision based on items with respect to a set of properties we deem important. To achieve this we need to  
\begin{itemize}
\item[($i$)] determine which properties of decision (Boolean) functions are important (Section \ref{sec:boolean-functions}),

\item[($ii$)] know how to decompose Boolean functions to evaluate the properties (Section \ref{sec:fourier-analysis}),

\item[($iii$)] adapt the Boolean decomposition to incorporate the dependence structure, which we do using graphical models (Section \ref{sec:p-biased-analysis}),

\item[($iv$)] create a measurement error model and connect it to classical and modern test theory (Sections \ref{sec:influence} and \ref{sec:measurement-error-classical}),

\item[($v$)] determine how the measurement error affects the Fourier analysis of Boolean functions (Section \ref{sec:stability-reliability}), and finally, 

\item[($vi$)] evaluate the properties of decision functions (Section \ref{sec:merit-sum-score})

\end{itemize}

Having obtained ($i$)-($vi$), we will argue in Section \ref{sec:merit-sum-score} that an extended form of a weighted sum score (see Section \ref{sec:boolean-functions}) seems the most appropriate way to aggregate items for making decisions, because a (weighted) sum score with threshold is the most stable decision function, and because the decision agrees most to the input when the decision is based on a (weighted) sum score with threshold. We illustrate some of the connections between the world of graphs, items and Fourier analysis of Boolean functions with simulations in Section \ref{sec:numerical-illustration}.

%We first define the mechanism of measurement error in Section \ref{sec:measurement-error-classical} and show that this definition complies with the ideas of classical test theory, as described in \citet{Lord:1968}. Then, in Section \ref{sec:stability-reliability} we define stability of the decision function and show how this is related to reliability of the items and to reliability of the decision function. We then use these concepts in Section \ref{sec:merit-sum-score} to argue that the sum score is the best decision function with respect to a set of desirable properties. We describe how to apply these ideas (and give statistical guarantees) by illustrating with numerical results in Section \ref{sec:numerical-illustration}. Proofs can be found in Appendix \ref{sec:proofs}. 

%---------------------------------------------------------------------------
\section{Boolean and decision functions}\label{sec:boolean-functions}\noindent
A Boolean function is a mapping from $\{-1,1\}^{n}$ of input configurations to the binary decision $\{-1,1\}$ \citep{Garban:2014}. Let the item $X_i$ be a random variable that takes values in the binary set $\{-1,1\}$. All items are collected in the $n$-vector $X\in \{-1,1\}^n$. We call the Boolean function $f:\{-1,1\}^n\to \{-1,1\}$ a {\em decision function} because of the binary classification corresponding to a two-choice decision. It is also possible to define a Boolean function with respect to the domain and range $\{0,1\}$ instead of $\{-1,1\}$, but we choose the latter because it is more convenient when defining the basis of the Fourier space and computing correlations. We collect all items in the set $V$, and we refer to items either with their index $i$ or with the random variable $X_i$.

\begin{example}\label{ex:majority-3}
An example of a decision function is the majority function on three items, where $X$ is the vector $(X_1,X_2,X_3)$. 
For three items there are $2^{3}=8$ patterns possible.
%%
%\begin{align*}
%&\{ (-1,-1,-1), (1,-1,-1),(-1,1,-1),(-1,-1,1),\\
%&\quad (1,1,-1),(1,-1,1),(-1,1,1),(1,1,1)\}
%\end{align*}
%
The majority function ${\rm maj}_{n}:\{-1,1\}^{n}\to \{-1,1\}$ is given by \citep{ODonnell:2014}
\begin{align}\label{eq:majority-rule-minus1} %{eq:majority-rule-zero-one}
\text{maj}_{n}(X)=
\begin{cases}
\phantom{-}1	&\text{ if } \sum_{i} X_{i} > 0\\
-1	&\text{ if } \sum_{i} X_{i} < 0\\
\end{cases}
\end{align}
For $n$ odd ${\rm maj}_{n}$ is positive if more item responses take the value $1$ than $-1$, and it is negative if the reverse holds. The values for each of the 8 patterns for any $X$ in $\{-1,1\}^{3}$ are
\begin{align}\label{eq:maj3-complete}
\begin{matrix}
&{\rm maj}_{3}(-1,-1,-1)=-1	&\quad{\rm maj}_{3}(-1,1,1)=+1\\
&{\rm maj}_{3}(-1,-1,1)=-1		&\quad{\rm maj}_{3}(1,-1,1)=+1\\
&{\rm maj}_{3}(-1,1,-1)=-1		&\quad{\rm maj}_{3}(1,1,-1)=+1\\
&{\rm maj}_{3}(1,-1,-1)=-1		&{\rm maj}_{3}(1,1,1)=+1
\end{matrix}
\end{align}
%

%
%\begin{align}\label{eq:majority-rule-minus1}
%{\rm maj}_{n}(X)={\rm sgn}(X_{1}+X_{2}+\cdots + X_{n})
%\end{align}
%
%where the function ${\rm sgn}:\mathbb{R}\to \{-1,1\}$ is the sign function and $\mathbb{R}$ is the set of real numbers. 
\end{example}

The majority function is an example of a class of functions called linear threshold functions (LTF), defined by functions $f:\{-1,1\}^{n}\to \{-1,1\}$ such that 
\begin{align}\label{eq:ltf}
f(X) = \text{sgn}(a_{0}+a_{1}X_{1}+\cdots + a_{n}X_{n})
\end{align}
where the function ${\rm sgn}(x)$ is the sign function and is $-1$ if $x<0$ and $1$ if $x>0$, and the $a_{i}$ are constants in $\mathbb{R}$ (the real line) and $a_{0}$ is referred to as the threshold \citep{ODonnell:2014}. The majority function on $\{-1,1\}^n$ in (\ref{eq:majority-rule-minus1}) is an example where $a_0=0$ and $a_i=1$ for $1\le i\le n$. LTFs are a more general way of thinking of the sum score. 
%Often the constants $a_{1},\dots, a_{n}$ are scaled such that $\sum a_{i}^{2}=1$, which is convenient when considering contrasts or the central limit theorem. 
Many types of item aggregation belong to the class of LTF. For example, grading is a form of LTF, where the threshold $a_{0}$ is determined according to a proportion expected to be answered correctly and $a_i$, with $1\le i\le n$, are weights for each item. Or a psychiatric diagnosis may be obtained for some empirically determined value $a_{0}$ and the $a_i$, with $1\le i\le n$, are weights that indicate relative importance to a disorder. But there are examples of items that will make the decision function necessarily nonlinear. For instance, in the situation of testing, item $i$ is classified as correct if and only if item $j$ is correct, then we obtain $x_{i}x_{j}$, and so $f$ is not a linear function. 
%Another example is a skip-item, where $x_{i-1}$ has the property that when it has value $-1$, then a set of subsequent items need not be answered, that is,
%
%\begin{align*}
%\begin{cases}
%x_{i}=x_{i+1}=\cdots = x_{i+m}=\varnothing	&\text{ if } x_{i-1}=-1\\
%x_{i}, x_{i+1}, \dots, x_{i+m} =\pm 1			&\text{ if } x_{i-1}=1
%\end{cases}
%\end{align*}
%
%Sometimes it is argued that all values for $x_{i}$ up to $x_{i+m}$ should be scored as $-1$, but in general this is a difficult issue with possibly large consequences for the decision. Even in the case where all items following the skip-item are coded as $-1$, then the influence (impact) of that item is often disproportional. 
%All input functions can be written as {\em polynomial threshold functions}, where a polynomial is said to have degree $k$ if the highest power in the terms of the polynomial is $k$. We will not discuss polynomial threshold functions here, see \citet{ODonnell:2014} for more discussion on this. 

\vspace{1em}\noindent
{\bf Properties of Boolean functions}\\\noindent
The majority function is a relatively simple function but turns out to have strong properties. It will turn out that the majority function is not very sensitive to small changes in the input; this will be one of our motivations to argue for linear threshold functions like the majority function (see more on this in Section \ref{sec:merit-sum-score}). But of course, it is in general reasonable to ask what kind of properties we would like our decision function to have so that we can interpret the results in a meaningful way. There are several properties that seem reasonable \citep{Kelly:1988}. A Boolean function $f$ is said to be\\

\begin{itemize}
\item[(a)] {\em monotone} or is {\em positively responsive} if for $X\le Y$ (i.e., $X_{j}\le Y_{j}$ for all $j$) implies that $f(X)\le f(Y)$;
\item[(b)] {\em odd} or {\em neutral} if $f(-X)=-f(Y)$;
\item[(c)] {\em unanimous} if $f(-1,-1,\ldots,-1)=-1$ and $f(1,1,\ldots,1)=1$;
\item[(d)] {\em symmetric} or {\em anonymous} if for any permutation $\text{perm}_n:\{-1,1\}^{n}\to \{-1,1\}^{n}$ of the coordinates in $X$ we have $f(X^{\text{perm}})=f(X)$;
\item[(e)] {\em transitive-symmetric} if for any $i\in V$ there is a permutation $\text{perm}_n:\{-1,1\}^{n}\to \{-1,1\}^{n}$ of the coordinates in $X$ that puts $X_{i}$ in place of $X_{j}$, such that $f(X^{\text{perm}})=f(X)$.\\
\end{itemize}
Monotonicity (a) is the requirement that the decision preserves the order of $X$ and $Y$, and so $f$ is always increasing (or decreasing). Equivalently, if $X$ has decision $f(X)=1$, and $f$ is monotone, then increasing some $X_j=-1$ to $1$, should result in the decision remaining $1$. This property is relevant in many contexts, like testing. Examples of monotone functions are the sum score $f(X)=\sum_i X_i$ and the dictator function $f(X)=X_i$, for some $i$, but the constant function that is always $-1$ for all input is not monotone \citep[see][for more examples]{Rosenbaum:1984}. We will see in later sections that monotonicity provides computational and theoretical advantages over non-monotone functions.  
%Monotonicity also stops trivial functions being of interest: If we consider the constant function $\text{const}_{1}$ that gives a $1$ no matter the input, we do not have monotonicity (e.g., $x=(1,1)$ and $y=(1,1)$. 
For an odd function in (b) we have that if all inputs were to be reversed then the decision is reversed. 
Unanimity in (c) requires that if all inputs are $-1$ (or $1$) then the decision has to be in the same direction (see also Section \ref{sec:rousseau}).
Symmetry or anonymity in (d) means that any decision $f(X)$ does not depend on where the item in the list was located; for any permutation of the items, the decision should be the same. Its weaker version, transitive-symmetry, in (e)  requires that coordinates $i$ and $j$ are exchangeable. There has to be a permutation for each pair $i$ and $j$ in $V$ such that the decision remains the same for the original and permuted version. It is easy to see that if $f(X^{\text{perm}})=f(X)$ for any permutation $\text{perm}_n$, then it holds for any subset of permutations that takes $i$ to coordinate $j$, and so (d) implies (e).

It has been shown that the majority function $\text{maj}_{n}$ in (\ref{eq:majority-rule-minus1}) has properties (a)-(e).
%---------------------------------------------------------------------------
%\setcounter{count}{0}
\begin{theorem}\label{lem:majority-properties}\citep{May:1952}
The majority function $\text{maj}_{n}:\{-1,1\}^{n}\to \{-1,1\}$ defined by $\text{maj}_{n}(x)=\text{sgn} \sum_{\i=1}^{n} x_{i}$ has the properties (a)-(e) above.
\end{theorem}
For completeness, we provide a proof in Appendix \ref{proof-sec:may-thm}. May's theorem also says that the only function that satisfies (a), (b) and (d) simultaneously is the majority function $\text{maj}_{n}$ \citep[][but see \citet{Kelly:1988} for an excellent discussion and proof]{May:1952}. So the majority function is an excellent candidate to use for decision making. But we believe these five properties are not sufficient, and should be augmented with an additional property. 

A property that we deem relevant for decision functions is stability. Stability refers to the idea that if a small proportion of items changes sign, then this will not immediately lead to a change in decision (which will be made more precise in Section \ref{sec:measurement-error-classical}). Stability is related to the idea of measurement error and reliability (Section \ref{sec:stability-reliability}). We suppose that there is some error in measurement for each item leading to the observed score $Y$ obtained from the original score $X$. We would want that the decision based on the decision function $f$ is not altered by changing a proportion $\tfrac{1}{2}(1-\rho)$ of $X$, where $\rho\in [-1,1]$.  A Boolean function $f$ is said to be\\

\begin{itemize}
\item[(f)] {\em stable} if $Y$ equals $X$ with probability $\tfrac{1}{2}(1+\rho)$ for each coordinate such that with high probabiilty probability $f(X)=f(Y)$. \\
\end{itemize}
%
%We consider stability in two versions. First, we consider stability upon possibly changing a single item; if $y$ is a copy of $x$ except for a single coordinate (with probability $\tfrac{1}{2}$), then we want to know the probability that $f(x)\ne f(y)$, so that the decision has changed. We call this influence of an item on the decision. 
%Each item in a test should have approximately the same influence on the decision. 
%We show that influence is related to the number of connections the item has in the graph representing conditional dependencies between items. 
These concepts will be made precise later. We will see (Section \ref{sec:measurement-error-classical}) that the correlation between the true and observed score from classical test theory \citep[see, e.g.,][]{Lord:1968} is proportional to the parameter $\rho$ in (f). Using the classical concept of reliability we will show that decision reliability between $f(X)$ and $f(Y)$, is co-determined by the reliability at the level of the items. We will also show that stability in the sense described above is directly related to reliability of the decision, and hence, is a reasonable property of a decision function. In order to obtain these results we require to decompose the decision function $f$ so that we can more easily work with the function. We next show that we can use the Fourier expansion of Boolean functions for this.

%---------------------------------------------------------------------------
\section{Fourier analysis of Boolean functions}\label{sec:fourier-analysis}\noindent
We aim to  compute the variance of a particular Boolean function $f:\{-1,1\}^{n}\to \{-1,1\}$ or the covariance with another Boolean function. In general, computing the variance or covariance on Boolean functions is not trivial, and hence, we require a method to decompose a Boolean function into more convenient components. We assume in this section that the variables are independent and have uniform measure; this is standard and the material presented in this section is mainly from \citet{ODonnell:2014}. We generalise these results to the case with non-uniform measure and dependence between variables in Section \ref{sec:p-biased-analysis}.

We begin with an example of a Fourier decomposition of a Boolean function. 

%\begin{example}
%We could base a decision on the maximum function $\max_{n}:\{-1,1\}^{n}\to \{-1,1\}$, which for $n=2$ is
%%
%\begin{align}\label{eq:max2-function-values}
%\begin{matrix}
%{\rm max}_{2}(-1,-1) = -1	&\quad {\rm max}_{2}(-1,1) = 1 \\
%{\rm max}_{2}(1,-1) = 1 &\quad{\rm max}_{2}(1,1) = 1\\
%\end{matrix}
%\end{align}
%%
%The Fourier expansion of ${\rm max}_{2}$ is 
%%
%\begin{align}\label{eq:max2-fourier}
%{\rm max}_{2}(X_{1},X_{2}) = \frac{1}{2} +\frac{1}{2}X_{1}+\frac{1}{2}X_{2}-\frac{1}{2}X_{1}X_{2}
%\end{align}
%%
%Basically this is an interpolation of the $\max_{2}$ function on the reals $\mathbb{R}$ where none of the variables is raised to a power (multilinear). The Fourier coefficients $\tfrac{1}{2}$ and $-\tfrac{1}{2}$ and the size of the sets of variables tell us something about the complexity of the function. 
%\end{example}
\begin{example}
We give the Fourier expansion of the majority function in (\ref{eq:majority-rule-minus1}) for independent variables $X=(X_1,X_2,X_3)$ each with probability $\mathbb{P}(X_i=1)=\tfrac{1}{2}$. The Fourier expansion of ${\rm maj}_{3}$  in (\ref{eq:majority-rule-minus1}) is
\begin{align}\label{eq:maj3-fourier}
{\rm maj}_{3}(X_{1},X_{2},X_{3}) = \frac{1}{2}X_{1}+\frac{1}{2}X_{2}+\frac{1}{2}X_{3}-\frac{1}{2}X_{1}X_{2}X_{3}
\end{align}
We will see how to obtain the coefficients below. 
\end{example}

To define the Fourier transform in general we define the parity function $\chi_{S}: \{-1,1\}^{k}\to \mathbb{R}$ for some set $S\subseteq V$ of size $|S|=k$
\begin{align}
\chi_{S}(X) = \prod_{i\in S}X_{i}
\end{align}
with the convention that for the empty set $S=\varnothing$ we have $\chi_{\varnothing}=1$. The parity function $\chi_{S}$ forms an orthonormal basis for the Fourier expansion for independent variables that have uniform measure \citep{ODonnell:2014}, similar to the sine and cosine functions for ordinary Fourier analysis. To define orthogonality we require an inner product. Here we use the inner product for two Boolean functions $f$ and $g$ defined by
\begin{align}\label{eq:inner-product}
\langle f,g\rangle = \frac{1}{2^{n}}\sum_{X\in \{-1,1\}^{n}}f(X)g(X)=\E_{\frac{1}{2}}(f(X)g(X))
\end{align}
where we used the $\tfrac{1}{2}$ in the expectation $\E_{\frac{1}{2}}$ to indicate that we assume that each variable $X_{j}$ is distributed uniformly and independently on $\{-1,1\}$. The uniform and independent distribution is not necessary but makes some ideas easier to explain. We extend all ideas to the general case where each item has its own probability and may be dependent on other items in Section \ref{sec:p-biased-analysis}. 

With the definition of the inner product we can define orthogonality by $\langle f,g \rangle=0$. Taking $f=\chi_{S}$ and $g=\chi_{T}$ as the functions in (\ref{eq:inner-product}) we obtain 
\begin{align}\label{eq:basis-function-orthogonal}
\langle \chi_{S},\chi_{T}\rangle=
\begin{cases}
1 	&\text{ if } S=T\\
0	&\text{ if } S\ne T
\end{cases}
\end{align}
\citep[For a proof, see][Theorem 1.5.]{ODonnell:2014} We can now define the Fourier expansion as follows. 
Let $f$ be a Boolean function and $\chi_{S}$ the parity function for subsets $S$ of $V$. Then the {\em Fourier expansion} is \citep[][Theorem 1.8]{ODonnell:2014}
\begin{align}
f(X) = \sum_{S\subseteq V} \hat{f}^{\frac{1}{2}}(S) \chi_{S}(X)
\end{align}
where $\hat{f}^{\frac{1}{2}}(S)$ is the {\em Fourier coefficient} on $S$ obtained with the uniform measure, defined for subset $S\subseteq V$ by
\begin{align}\label{eq:fourier-coefficients}
\hat{f}^{\frac{1}{2}}(S)=\langle f, \chi_{S}\rangle = \E_{\frac{1}{2}}(f(X)\chi_{S}(X))
\end{align}
Note that for $S=\varnothing$ we have $\hat{f}^{\frac{1}{2}}(\varnothing)=\langle f,1\rangle = \E_{\frac{1}{2}}(f(X))$.

%\begin{example}
%We return to the example of the function ${\rm max}_{2}$ to compute the Fourier coefficients using (\ref{eq:fourier-coefficients}). For each subset $S$ of $V$ we require the expectation of ${\rm max}_{2}(X)\chi_{S}$. For $S=\varnothing$ we have that $\E_{\frac{1}{2}}({\rm max}_{2}(X)\chi_{\varnothing}(X)) = \E_{\frac{1}{2}}({\rm max}_{2}(X))$, and so using (\ref{eq:max2-function-values})
%%
%\begin{align*}
%\widehat{\rm max}_{2}(\varnothing) &= \Prob_{\frac{1}{2}}({\rm max}_{2}(X)=1) -  \Prob_{\frac{1}{2}}({\rm max}_{2}(X)=-1)=\frac{3}{4}-\frac{1}{4}=\frac{1}{2}
%\end{align*}
%% 
%From (\ref{eq:max2-function-values}) we obtain that for the subsets $S=\{1\}$ or $\{2\}$, we have that ${\rm max}_{2}(X)X_{1}$ equals 1 in 3 out of 4 cases and so
%%
%\begin{align*}
%\widehat{\rm max}_{2}(\{1\})&=\frac{3}{4}-\frac{1}{4}=\frac{1}{2}
%\end{align*}
%% 
%for both sets $\{1\}$ and $\{2\}$. Then, for $S=\{1,2\}$ we obtain $\max_{2}(X)X_{1}X_{2}$ equals 1 in 1 out of 4 cases, and so
%%
%\begin{align*}
%\widehat{\rm max}_{2}(\{1,2\})&=\frac{1}{4}-\frac{3}{4}=-\frac{1}{2}
%\end{align*}
%% 
%These are exactly the coefficients we had in the Fourier expansion in (\ref{eq:max2-fourier}).
%\end{example}

%
\begin{example}
The Fourier coefficients for the function ${\rm maj}_{3}$ function in (\ref{eq:majority-rule-minus1}) can be obtained using the relation $\widehat{{\rm maj}}_{3}(S)= \E_{\frac{1}{2}}({\rm maj}_{3}(X)\chi_{S}(X))$ for each $S$. With $S=\varnothing$ we get by inspection of (\ref{eq:maj3-complete})
\begin{align*}
\mathbb{E}_{\frac{1}{2}}(\text{maj}_3\cdot 1)=4\frac{1}{2} - 4\frac{1}{2} = 0
\end{align*}
and for $S=\{1\}$ we obtain, again by inspection of (\ref{eq:maj3-complete}) 
\begin{align*}
\mathbb{E}_{\frac{1}{2}}(\text{maj}_3\cdot X_1)=3\frac{1}{2} -\frac{1}{2}  = \frac{1}{2}
\end{align*}
For all 8 possible values of $\text{maj}_3$ we get the coefficients
%
%%
%\begin{table}
%\begin{tabular}{l c c c c}
%$S$		&$\varnothing$		&$\{1\}, \{2\}, \{3\}$		&$\{1,2\}, \{1,3\}, \{2,3\}$		&$\{1,2,3\}$\\ \midrule
%$\widehat{{\rm maj}}_{3}(S)$
%		&$\E({\rm maj}_{3})$	&$\E({\rm maj}_{3}\chi_{\{1\}})$	&$\E({\rm maj}_{3}\chi_{\{1,2\}})$	&$\E({\rm maj}_{3}\chi_{\{1,2,3\}})$\\[.5em]
%		&0				&$\frac{1}{2}$			&0						&$-\frac{1}{2}$\\
%		\midrule
%\end{tabular}
%\caption{Fourier coefficients of the majority function with $n=3$. }
%\label{tab:maj3-coefficients}
%\end{table}
%
\begin{align*}
\begin{matrix}
\widehat{\text{maj}}_{3}(\varnothing) =& 0\qquad			&\widehat{\text{maj}}_{3}(\{1,2\}) =& 0\\
\widehat{\text{maj}}_{3}(\{1\}) = &\frac{1}{2}\qquad		&\widehat{\text{maj}}_{3}(\{1,3\}) = &0\\
\widehat{\text{maj}}_{3}(\{2\}) = &\frac{1}{2}\qquad		&\widehat{\text{maj}}_{3}(\{2,3\}) = &0\\
\widehat{\text{maj}}_{3}(\{3\}) = &\frac{1}{2}\qquad		&\widehat{\text{maj}}_{3}(\{1,2,3\}) = &-\frac{1}{2}
\end{matrix}
\end{align*}
It is easily seen that the coefficients correspond to those in (\ref{eq:maj3-fourier}).
\end{example}

The Fourier expansion of a Boolean function makes it easier to determine the mean, variance, and covariance of Boolean functions. An often used relation in this context, known as Plancheral's theorem, is that for any Boolean functions $f,g$
\begin{align}\label{eq:plancheral}
\langle f, g\rangle = \E_{\frac{1}{2}}(f(X)g(X)) = \sum_{S\subseteq V} \hat{f}^{\frac{1}{2}}(S)\hat{g}^{\frac{1}{2}}(S)
\end{align}
From this we can characterise the mean and variance in terms of Fourier coefficients. 
If we set $g=1$, we already noted that in a Fourier context we select $S=\varnothing$ for $\chi_{S}$, and we get
\begin{align}
\langle f,1\rangle = \E_{\frac{1}{2}}(f(X))=\hat{f}^{\frac{1}{2}}(\varnothing)
\end{align}
And for the variance we obtain (a variant of Parseval's Theorem)
\begin{align}\label{eq:parseval}
\text{var}_{\frac{1}{2}}(f)=\langle f-\E_{\frac{1}{2}}(f(X)),f-\E_{\frac{1}{2}}(f(X))\rangle = \sum_{S\neq\varnothing}\hat{f}^{\frac{1}{2}}(S)^{2}
\end{align}
where from the Fourier sum the mean $\hat{f}^{\frac{1}{2}}(\varnothing)=\E_{\frac{1}{2}}(f(X))$ has been removed. The covariance between Boolean functions $f$ and $g$ is then
\begin{align}
\text{cov}_{\frac{1}{2}}(f,g)=\langle f-\E_{\frac{1}{2}}(f(X)),g-\E_{\frac{1}{2}}(g(X))\rangle = \sum_{S\neq\varnothing}\hat{f}^{\frac{1}{2}}(S)\hat{g}^{\frac{1}{2}}(S)
\end{align}
We use these identities often, and especially when determining the effects of inducing measurement error in Section \ref{sec:stability-reliability}.

%
%---------------------------------------------------------------------------
\section{General Fourier analysis of Boolean functions: Capturing dependence and bias with graphical models}\label{sec:p-biased-analysis}\noindent
For the Fourier analysis it is common to assume that 
\begin{itemize}
\item[(1)] for each item $i\in V$ the probability is uniform over $-1$ and $1$, i.e., $\mathbb{P}_{\frac{1}{2}}(X_i=1)=\mathbb{P}_{\frac{1}{2}}(X_i=-1)=\tfrac{1}{2}$, and 

\item[(2)] for all $i\in V$ the $X_i$ are independent, i.e., $\mathbb{P}_{\frac{1}{2}}(X_1 X_2)= \mathbb{P}_{\frac{1}{2}}(X_1)\mathbb{P}_{\frac{1}{2}}(X_2)$.

\end{itemize}
From (1), assuming the uniform measure over $\{-1,1\}$, we obtain $\E_{\frac{1}{2}}(X_{i})=-1\cdot\tfrac{1}{2} + 1\cdot\tfrac{1}{2}=0$ for all $i\in V$. We call this unbiased. Obviously, this is unrealistic and we need to accommodate the $p$-biased situation where $\mathbb{E}(X_i) \ne 0$. From (2) we get the result that $\mathbb{E}_{\frac{1}{2}}(X_1X_2)=\mathbb{E}_{\frac{1}{2}}(X_1)\mathbb{E}_{\frac{1}{2}}(X_2)$; we required this to obtain the orthonormal basis (parity functions) in (\ref{eq:basis-function-orthogonal}). When the $X_i$ are dependent, this separation is not possible in general, and so we cannot obtain an orthonormal basis as before. 
To overcome these two assumptions, we introduce a formalism using graphical models that achieves an orthonormal basis and can accommodate the $p$-biased case.

\vspace{1em}\noindent
{\bf Graphical models}\\
Let $G=(V,E)$ be an undirected graph, where $V=\{1,2,\ldots,n\}$ is the set of nodes and $E\subseteq V\times V$ is the set of edges $\{(i,j): i,j\in V\}$, with size $|E|$. Nodes that are connected are called adjacent or neighbours. Let $\partial i$ be the set of nodes that are adjacent to (neighbours of) node $i$, $\{j\in V\backslash \{i\} : (i,j)\in E\}$. 
For a set of nodes $B$ we denote its adjacent nodes $\{j\in V\backslash B: (i,j)\in E \text{ and } i\in B\}$ by $\partial B$; the set $\partial B$ is also referred to as the {\em boundary} of $B$. A subset of nodes $W$ is a cutset or separator set of the graph if removing $W$ results in two (or more) components. For instance, $W$ is a cutset if any path between any two nodes $s\in A$ and $t\in B$ must go through some $q\in W$. A clique is a subset of nodes in $C\subset V$ such that all nodes in $C$ are connected, that is, for any $i,j\in C$ it holds that $(i,j)\in E$. A maximal clique is a clique such that including any other node in $V\backslash C$ will not be a clique. 

\begin{figure}[t]
\begin{minipage}[b]{0.45\textwidth}\centering
\begin{align*}
\xymatrix{
*++[o][F]{4}\ar@{}[r]_>{\displaystyle C_{2}}&{\xy(-19,-20);(-6,5) **\frm<44pt>{--}\endxy}	&*++[o][F]{1}\ar@{-}[rr]\ar@{-}[d]\ar@{-}[r]	&&*++[o][F]{2}\\
*++[o][F]{5}\ar@{-}[u]^{\displaystyle C_{3}} 	&{\xy(4,-5);(45,20) **\frm<44pt>{--}\endxy} &*++[o][F]{3}\ar@{-}[urr]\ar@{-}[ull] &
\xy 
/r6pc/:p-(.85,-0.32),
%{\ellipse<,14pt>{-}}
{\ellipse(,.31){--}}
\endxy
&C_{1}
}
\end{align*}
(a)
\end{minipage}
\hspace{2em}
\begin{minipage}[b]{0.45\textwidth}\centering
\begin{align*}
\xymatrix{
*++[o][F**:pink]{4}\ar@{}[r]			&	&*++[o][F]{1}\ar@{-}[rr]\ar@{-}[d]\ar@{-}[r]	&&*++[o][F]{2}\\
*++[o][F]{5}\ar@{-}[u]		&{\xy(-19,-6);(-6,5) **\frm<44pt>{--}\endxy}{\xy(4,-5);(45,20) **\frm<44pt>{--}\endxy} 	&*++[o][F]{3}\ar@{-}[urr]\ar@{-}[ull]&&
}
\end{align*}
(b)
\end{minipage}
\caption{Graph of 5 nodes with cliques $C_{1}=\{1,2,3\}$, $C_{2}=\{3,4\}$ and $C_{3}=\{4,5\}$. In (a) the factorisation property obtained from the three cliques, such that the distribution is defined for potential functions $\psi_{C_{1}}$, $\psi_{C_{2}}$, and $\psi_{C_{3}}$, as in (\ref{eq:factorisation-property}). In (b) conditional independence is shown for $X_{5}\independent \{X_{1},X_{2},X_{3}\}\mid X_{4}$. }
\label{fig:graphical-model}
\end{figure}
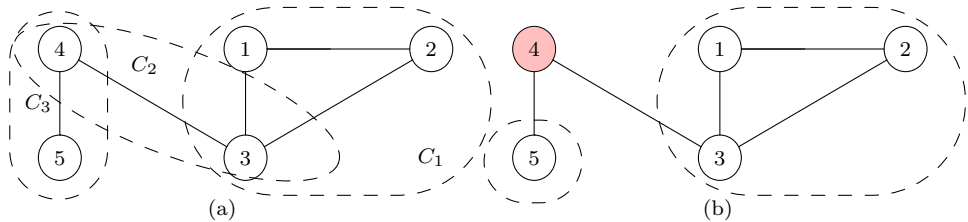

\begin{example}
Consider the example graph in Figure \ref{fig:graphical-model}(a). There are 5 nodes and three cliques $C_{1}=\{1,2,3\}$, $C_{2}=\{3,4\}$ and $C_{3}=\{4,5\}$. The clique $C_{2}$ is a cutset because removing $C_{2}$ will result in two components. Equivalently, we see that all paths from clique $C_{1}$ to $C_{3}$ go through $C_{2}$.
\end{example}

For an undirected graph $G$, we associate with each node $i\in V$ a random variable $X_{i}$ over the set  $\{-1,1\}$. For any subset $A\subset V$ of nodes we define a configuration $X_{A}=\{X_{i}: i\in A\}$. Two variables $X_{i}$ and $X_{j}$ are independent if $\Prob(X_{i},X_{j})=\Prob(X_{i})\Prob(X_{j})$, and we write this as $X_{i}\independent X_{j}$. The variables $X_{i}$ and $X_{j}$ are conditionally independent on $X_{k}$ if $\Prob(X_{i},X_{j}\mid X_{k})=\Prob(X_{i}\mid X_{k})\Prob(X_{j}\mid X_{k})$.
For subsets of nodes $A$, $B$, and $W$, we denote by $X_{A}\independent X_{B}\mid X_{W}$ that $X_{A}$ is conditionally independent of $X_{B}$ given $X_{W}$. 

A random vector $X$ is {\em Markov with respect to $G$} if 
\begin{align}\label{eq:markov-property}
\text{$W$ is a cutset for disjoint subsets $A$ and $B$} \quad\implies\quad X_{A}\independent X_{B}\mid X_{W}
\end{align}
An equivalent way to define a Markov random field is in terms of the factorisation over cliques of the distribution function, which reveals conditional independence between cliques. For each clique $C$ in the set of all cliques $\mathcal{C}$  of graph $G$ a compatibility function $\psi_C: \{0,1\}^{n} \to \mathbb{R}_{+}$ maps the states of the nodes in clique $C$ to the positive reals. When normalized, the product of the compatibility functions defines the distribution. 
The distribution of the random vector $X$ {\em factorises according to graph $G$} if it can be represented by a product of compatibility functions of the cliques
\begin{align}\label{eq:factorisation-property}
\Prob(X=x) = \frac{1}{Z}\prod_{C\in \mathcal{C}} \psi_{C}(x_{C})
\end{align}
where $Z$ is the normalisation constant. Here we use the functions $\psi_{C}(x_{C})=\exp(f(x_{C}))$ for exponential family models. For strictly positive distributions the Hammersly-Clifford theorem says that the Markov and factorisation properties are equivalent \citep{Cowell:1999,Lauritzen96,Wainwright:2019}.  

%\xy 
%/r6pc/:p-(0.3,-2),
%{\ellipse(,.05){--}}
%\endxy
%

\begin{example}\label{ex:clique-set}
Considering Figure \ref{fig:graphical-model}(a) again, we see that the factorisation over cliques is 
\begin{align*}
\psi_{C_{1}}(x_{1},x_{2},x_{3})\psi_{C_{2}}(x_{3},x_{4})\psi_{C_{3}}(x_{4},x_{5})
\end{align*}
And, in Figure \ref{fig:graphical-model}(b) the Markov property implies that $X_{5}\independent X_{3}\mid X_{4}$, for instance. Such conditional independence can be checked visually from the paths that go through node 4 to get from 5 to 3. Normalising each of the clique functions $\psi_{C_i}$ then gives a conditional product measure
\begin{align*}
\mathbb{P}_{C_{1}}(x_{1},x_{2}\mid x_{3})\mathbb{P}_{C_{2}}(x_{4}\mid x_3)\mathbb{P}_{C_{3}}(x_{5},x_{4})
\end{align*}
\end{example}

The Markov and factorisation properties imply that in an undirected graphical model each node is conditionally independent of other nodes given the nodes adjacent to it. That is, for any disjoint subsets $B$ and $D\backslash \partial B\subset V$, where the boundary $\partial B$ is excluded, we have the conditional independence $X_{B}\independent X_{D\backslash \partial B}\mid X_{\partial B}$. So, there is in general conditional independence for nodes when conditioned on their boundary sets. It is this conditional independence that we use to obtain the a conditional product measure. 

%

%-------------------------------------------------------
\vspace{1em}\noindent
{\bf Pseudo-likelihood approximation for product measure}\label{sec:general-fourier-analysis}\\ \noindent
If we have conditional independence, we can obtain a similar construction as if there was independence, that is, $\E(X_{1}X_{2}\mid X_{3})=\E(X_{1}\mid X_{3})\E(X_{2}\mid X_{3})$, if $X_{1}$ is conditionally independent of $X_{2}$ given $X_{3}$. However, whenever there is conditional dependence between variables in the subset $S$ (e.g., clique $C_1$ in Example \ref{ex:clique-set}), then we cannot use the product of expectations. We therefore propose to work with the pseudo-likelihood, the product of conditional distributions \citep{Besag:1974,Hyvarinen:2006}
\begin{align}\label{eq:joint-measure}
\Prob_{\pi}(x):=\Prob(x_{1}\mid x_{\partial 1})\Prob(x_{2}\mid x_{\partial 2})\cdots \Prob(x_{n}\mid x_{\partial n})
\end{align}
We write $\Prob_{\pi}$ for the measure in (\ref{eq:joint-measure}), with probabilities $\pi=(p_{1},p_{2},\ldots,p_{n})$ and $p_i=\mathbb{P}_{p_i}=\Prob(X_{i}=1\mid x_{\partial i})$; we write $\partial i$ for the boundary set (Markov blanket) of variable $X_i$.

Here we use the pseudo-likelihood as an approximation to the joint distribution. It has been shown that parameter estimation with the pseudo-likelihood obtains consistent estimators \citep{Hyvarinen:2006,Nguyen:2017}.  A special case that we use here is for pairwise models, specifically the Ising model defined on graph $G$ \citep{Cipra:1987,Wainwright:2008}, or, as it is sometimes called, the auto-logisitc model \citep{Besag:1974}, which has been shown to be consistent  \citep[see, e.g., ][]{Loh:2013,Yang:2012,Yang:2013,Haslbeck:2015}. 
Let $X$ be a random vector over $\{-1,1\}^{n}$. The joint distribution of $X$ is called the {\em Ising model} if its probability distribution is defined as 
\begin{align}\label{eq:ising-joint-probability}
\Prob(x) = \frac{1}{Z}\exp\left( \sum_{i\in V}\xi_{i}x_{i} + \sum_{(i,j)\in E}\theta_{ij} x_{i}x_{j}\right)
\end{align}
where
\begin{align}\label{eq:normalising-constant}
Z_{V}=\sum_{x\in \{-1,1\}^{n}} \exp\left( \sum_{i\in V}\xi_{i}x_{i} + \sum_{(i,j)\in E}\theta_{ij} x_{i}x_{j}\right)
\end{align}
And the conditional distribution of $X_i$ given the Markov blanket $X_{\partial i}$ is 
\begin{align}\label{eq:ising-conditional-probability}
\Prob_{p_i}(x_{i}\mid x_{\partial i}) = 
\frac{\exp\left(x_{i}  \left(\xi_i + \frac{1}{2}\sum_{j\in \partial i}\theta_{ij} x_{j}\right)\right)}
	{\exp\left(-  \left(\xi_i + \frac{1}{2}\sum_{j\in \partial i}\theta_{ij} x_{j}\right)\right)+\exp\left(\xi_{i} + \frac{1}{2}\sum_{j\in \partial i}\theta_{ij} x_{j}\right)}
\end{align}
The factor $\tfrac{1}{2}$ is included because we combine all conditionals to form the pseudo-likelihood, and in this combination each term occurs twice (see Lemma \label{lem:conditional-product-measure}\ref{lem:conditional-product-measure} in Appendix \ref{proof-sec:conditional-product-measure}). The conditional distribution in (\ref{eq:ising-conditional-probability}) has been used to obtain estimates of the parameters, since using the conditional distribution removes the problem of having to compute the normalisation constant for the joint probability \citep{Wainwright:2008,Waldorp:2019}. 

We are consistent with the univariate conditional probabilities to the joint distribution in the sense that we only require a rescaling. 
We can express the difference between the joint $\Prob$ and product of conditionals $\Prob_{\pi}$ in terms of the Kullback-Leibler divergence \citep{Cover:2006}. 

%---------------------------------------------------------------------------
\begin{lemma}\label{lem:kl-divergence-probability}
Let $\Prob$ be the joint Ising probability (\ref{eq:ising-joint-probability}) and $\Prob_{\pi}$ be the product of conditionals (\ref{eq:joint-measure}). Then the Kullback-Leibler divergence is
\begin{align*}
D_{\rm KL}(\Prob \mid\mid \Prob_{\pi})=\log \frac{Z_{V}}{\prod_{i\in V}Z_{i}}
\end{align*}
\end{lemma}
A proof is in Appendix \ref{proof-sec:kl-divergence}. This implies that by considering the conditional product measure (\ref{eq:joint-measure}), we are in fact also targeting the joint probability; the difference is only in a scaling factor.

\vspace{1em}\noindent
{\bf Orthonormal basis for dependent variables}\\\noindent
Using the product of conditionals $\Prob_\pi$ in (\ref{eq:joint-measure}) allows us to obtain the requirements of orthonormality similar to (\ref{eq:basis-function-orthogonal}) for the Fourier analysis but then for $p$-biased and dependent variables. The expectation of the basis function needs to be 0 and the expectation of the square needs to be 1. We therefore  define the function $\phi: \{-1,1\}\to \mathbb{R}$ by
\begin{align}\label{eq:z-transform-phi}
\phi(X_{i}) = \frac{X_{i}-\mu_{i}}{\sigma_{i}}
\end{align}
where we use the conditional expectation
\begin{align}\label{eq:p-biased-mean}
\mu_{i} = \E(X_{i}\mid X_{\partial i}) = 2p_{i}-1
\end{align}
and conditional variance
\begin{align}\label{eq:p-biased-variance}
\sigma_{i}^{2} = \E ((X_{i}-\mu_{i})^{2}\mid X_{\partial i}) = 4p_{i} (1-p_{i})
\end{align}
Note that $\phi(1)=\sqrt{(1-p_{i})/p_{i}}$ and $\phi(-1)=-\sqrt{p_{i}/(1-p_{i})}$. We write $\E_{p_{i}}$ for the conditional expectation $\E(\cdot\mid x_{\partial i})$ based on the conditional measure $\mathbb{P}_{p_i}:=\mathbb{P}(\cdot\mid x_{\partial i})$. For each $i\in V$ we then have that the mean is
$\E_{p_{i}}(\phi(X_{i}))=0$
and the variance is
$\text{var}_{p_{i}}(\phi(X_{i}))=1$. 
For the product function $\phi_{S}:\{-1,1\}^{n}\to \mathbb{R}$ defined by $\phi_{S}(X)=\prod_{i\in S}\phi(X_{i})$ for any $S\subseteq V$, we obtain with $\Prob_{\pi}$ in (\ref{eq:joint-measure})
\begin{align}\label{eq:expectation-product}
\E_{\pi}(\phi_{S}(X)) := \E_{p_{1}}(\phi(X_{1})) \E_{p_{2}}(\phi(X_{2}))\cdots \E_{p_{k}}(\phi(X_{k}))
\end{align}
where $k=|S|$.
This again results in orthogonality as before in (\ref{eq:basis-function-orthogonal}), but now for the functions $\phi_S$ with dependent $X_i$. 
\begin{proposition}($p$-biased and dependent orthonormal base) \label{lem:p-biased-basis}
Let $\phi_S(X)$ be the transformation in (\ref{eq:z-transform-phi})  for $X_i$ $p$-biased for $i$ and dependent. Using the pseudo-likelihood approach (\ref{eq:joint-measure}), we obtain for subset $S$ and $T$ of $V$
\begin{align}\label{eq:basis-function-orthogonal-p-biased}
\langle \phi_{S},\phi_{T}\rangle=
\E_{\pi}(\phi_{S}(X)\phi_{T}(X)) 
=
\begin{cases}
1	&\text{ if } S=T\\
0	&\text{ if } S\ne T\\ 
\end{cases}
\end{align}
\end{proposition}
A proof of this result can be found in Appendix \ref{proof-sec:p-biased-basis}. With this $p$-biased basis function for dependent variables, the $p$-biased Fourier expansion is
\begin{align}\label{eq:p-biased-fourier-coefficient}
f(x) = \sum_{S\subseteq V} \hat{f}^{\pi}(S)\phi_{S}(x)
\end{align}
where the Fourier coefficient is 
\begin{align}\label{eq:p-biased-coefficient}
\hat{f}^{\pi}(S)=\E_{\pi}(f(X)\phi_{S}(X))
\end{align}
and $\pi=(p_{1},p_{2},\ldots,, p_{n})$. 
To obtain the $p$-biased Fourier expansion we need only plug-in the transformed variables $X_{i}=\mu_{i}+\sigma_{i}\phi(X_{i})$. 

%\begin{example}
%Consider again the majority function on three items $\text{maj}_{3}$ with Fourier expansion under the uniform measure given in Example \ref{ex:majority-3}. Plugging in the transformed values for $x_{i}$ gives a $p$-biased Fourier expansion with Fourier coeffcients for the basis functions $\chi^\phi_{S}$ with $S\subseteq V$
%%
%\begin{align}\label{eq:maj3-p-biased-fourier}
%\begin{matrix}
%\widehat{\text{maj}}_{3}^{\pi}(\varnothing)= &\frac{1}{2}(\mu_{1}+\mu_{2}+\mu_{3}-\mu_{1}\mu_{2}\mu_{3}) 
%	&\widehat{\text{maj}}_{3}^{\pi}(\{1,2\})=&-\frac{1}{2}\mu_{3}\sigma_{1}\sigma_{2}\\
%\widehat{\text{maj}}_{3}^{\pi}(\{1\})=&\frac{1}{2}\sigma_{1}(1-\mu_{2}\mu_{3})
%	&\widehat{\text{maj}}_{3}^{\pi}(\{1,3\})=&-\frac{1}{2}\mu_{2}\sigma_{1}\sigma_{3}\\
%\widehat{\text{maj}}_{3}^{\pi}(\{2\})=&\frac{1}{2}\sigma_{2}(1-\mu_{1}\mu_{3})
%	&\widehat{\text{maj}}_{3}^{\pi}(\{2,3\})=&-\frac{1}{2}\mu_{1}\sigma_{2}\sigma_{3}\\
%\widehat{\text{maj}}_{3}^{\pi}(\{3\})=&\frac{1}{2}\sigma_{3}(1-\mu_{1}\mu_{2})
%	&\widehat{\text{maj}}_{3}^{\pi}(\{1,2,3\})=&-\frac{1}{2}\sigma_{1}\sigma_{2}\sigma_{3}
%\end{matrix}
%\end{align}
%%
%where $\pi=(p_{1},p_{2},p_{3})$ is used to indicate that they are $p$-biased coefficients. 
%%Note that under the uniform measure the coefficients for products with two variables $X_{i}X_{j}$ were all 0, whereas in the transformed version there are non-zero coefficients for $\phi(X_{i})\phi(X_{j})$.
%\end{example}

Using the $p$-biased case above and the dependence between $X_i$ with the pseudo-likelihood probability in (\ref{eq:joint-measure}), the mean and (co)variance of $f$ can be determined analogously to the unbiased and independent case. 

\begin{proposition}\label{lem:p-biased-mean-variance-covariance}
Let $f$ be a Boolean function and let the $X_{i}$ be biased and dependent in that $\E_{p_{i}}(X_{i})=2p_{i}-1$ for each $i$, based on the conditional distributions obtained with the Ising model (\ref{eq:ising-conditional-probability}). Then the mean is
\begin{align*}
\E_{\pi} (f(X)) &= \E_{\pi}\left( \sum_{S = \varnothing}\hat{f}^{\pi}(S)\phi_{S}(X)\right) =  \hat{f}^{\pi}(\varnothing)
\end{align*}
and the covariance is 
%%
%\begin{align*}
%\text{var}_{\pi}(f) = \E_{\pi} f(X)^{2} -(\E_{\pi} f(X))^{2}=  \sum_{S\neq\varnothing}\hat{f}^{\pi}(S)^{2}
%\end{align*}
%%
%and
%
\begin{align*}
\text{cov}_{\pi}(f) = \E_{\pi} f(X)g(X) -\E_{\pi} f(X)\E_{\pi} g(X)=  \sum_{S\neq\varnothing}\hat{f}^{\pi}(S)\hat{g}^{\pi}(S)
\end{align*}
The variance is obtained with $f=g$. 
\end{proposition}
For a proof see  Appendix \ref{p-biased moments}. 

We now have obtained a framework where we can start to investigate the effects of items on decisions using Fourier analysis of Boolean functions.

%Note that
%%
%\begin{align}\label{eq:mean-f}
%\E_{\pi} (f(X)) = \E_{\pi}\left( \sum_{S\subseteq V}\hat{f}^{\pi}(S)\chi^\phi_{S}(x)\right) =  \hat{f}^{\pi}(\varnothing)
%\end{align}
%%
%since $\E_{\pi}(\chi^\phi_{S}(x)) = 1$ only for the set $S=\varnothing$ and 0 otherwise. So
%%
%\begin{align}\label{eq:variance-f-parseval}
%\text{var}_{\pi}(f) = \E_{\pi} (f(X)^{2}) -(\E_{\pi} f(X))^{2}=  \sum_{S\neq\varnothing}\hat{f}^{\pi}(S)^{2}
%\end{align}
%%
%which is similar to the unbiased case (\ref{eq:parseval}) except that we plug in the biased Fourier coefficients $\hat{f}^{\pi}$ for dependent variables (see Lemma \ref{lem:p-biased-mean-variance-covariance} in the Appendix \ref{sec:proofs}). This version of Parseval's theorem implies that the variance of a Boolean function with outcome in $\{-1,1\}$ and expectation 0, is 1.

 %---------------------------------------------------------------------------
\section{Influence of items}\label{sec:influence}\noindent
To begin the analysis of the effect of measurement error, we investigate the effect of a single item on the decision, i.e., whether $f(X)$ changes if a single item in $X$ were flipped. This is called the influence. If the influence is disproportionately large, then this could raise a flag if it is not the intention of the test to have a single item that is necessary to pass. We obtain a bound of the influence, given certain conditions, using the Fourier coefficients, which is relatively simple for monotone functions, i.e., functions such that $X\le Y$ coordinatewise implies $f(X)\le f(Y)$ (see property (a) in Section \ref{sec:boolean-functions}). 

Let $X^i$ denote the vector $(X_1,\ldots, X_{i-1}, -X_{i},X_{i+1},\ldots,X_n)$, where we flip the sign of $X_i$ and leave the remaining elements in tact.
To get the probability for when the decision changes under $f$ when going from $X$ to $X^i$, i.e., $f(X)\ne f(X^{i})$, we obtain an indicator function $\mathbbm{1}\{f(X)\ne f(X^{i})\}$, and its expectation will give the probability. Such an indicator can be obtained from the discrete derivative function
\begin{align}
D_{i} f =\frac{1}{2}(f(X)-f(X^i))
\end{align}
Then we see that if the decision changes, then $D_{i}f$ is $-1$ or $1$, and if the decision remains the same, then $D_{i}f$ is 0. This means that $(D_{i}f)^{2}$ is the indicator function for whenever item $i$ results in $f(X)\ne f(X^{i})$.  And so
\begin{align*}
\Prob_\pi(f(X)\ne f(X^i))=\E_{\pi}(\mathbbm{1}\{f(X)\ne f(X^{i})\})=\E_{\pi}((D_{i}f(X))^{2})
\end{align*}
The discrete derivative of the function $\phi_{S}$ in terms of the Fourier coefficient is
\begin{align*}
D_{i} \phi_{S}(X) = \frac{1}{2\sigma_{i}}(1-\mu_{i}-(-1-\mu_{i}))\prod_{j\in S\backslash \{i\}} \phi(X_{j})=\frac{1}{\sigma_{i}}\prod_{j\in S\backslash \{i\}} \phi(X_{j})
\end{align*}
if $i\in S$ and 0 otherwise. And so, since $\E_{\pi}(\phi_{S\backslash \{i\}}(X)^{2})=1$ for each $S$ (see Lemma 
\label{lem:p-biased-influence}\ref{lem:p-biased-influence} 
in Appendix \ref{proof-sec:p-biased-influence})
\begin{align}
\mathbb{I}^{\pi}_{i}(f)=\E_{\pi}((D_{i}f(X))^{2})=\frac{1}{\sigma_{i}^{2}}\sum_{S\ni i}\hat{f}^{\pi}(S)^{2}
\end{align}
where $S\ni i$ means the sum across all subsets $S$ that contain $i$. 
For monotone Boolean functions like ${\rm maj}_{n}$, the influence can be simplified to the Fourier coefficient of the singleton set, that is (see Lemma %\label{lem:influence-fourier-s1}
6
%\ref{lem:influence-fourier-s1} 
in Appendix \ref{proof-sec:influence-fourier-s1})
\begin{align}\label{eq:inluence-fourier-monotone}
\mathbb{I}^{\pi}_{i}(f)=\frac{1}{\sigma_{i}}\hat{f}^{\pi}(i)
\end{align}
where $\hat{f}^{\pi}(i)=\hat{f}^{\pi}(\{i\})=\E_{\pi}(f(X)\phi(X_{i}))$. Using this relation, we can easily obtain an upper bound on the influence of an item. 

\begin{proposition}(Bound on influence)\label{prop:bound-influence}
Let $f$ be a Boolean monotone function and assume transitive symmetry (e) from Section \ref{sec:boolean-functions} for $p$-biased and dependent $X_i$ given the measure $\mathbb{P}_\pi$ in (\ref{eq:joint-measure}). Then 
\begin{align}\label{eq:bound-influence}
\mathbb{I}_{i}^{\pi}(f)\le \frac{\text{sd}_{\pi}(f)}{\sigma_{i}\sqrt{n}}
\end{align}
\end{proposition}
A proof can be found in Appendix \ref{proof-sec:bound-influence}. 
We could use this bound to see whether any of the items come close to this bound, suggesting that its influence maybe too large compared to the other items. 

An example of a test where it holds that each item is bounded in influence, is a balanced test. In a balanced test each item has roughly the same influence on the decision; a balanced test is then a stronger property than one where the influence is bounded. A balanced test can be visualised in terms of a graph, where the nodes are identified with the random variables (items) and the edges are conditional associations. An interesting result for balanced tests is that a graph with nodes that have the same degree and the same coefficients in terms of the Ising model (\ref{eq:ising-joint-probability}), has equal influences for each of the nodes in the graph. 

\begin{proposition}(Equal degree implies equal influence)\label{prop:influence-same-degree-same}
Let the graph $G$ be induced by a regular (equal degree nodes) Ising model (\ref{eq:ising-joint-probability}) with equal interaction and threshold parameters. Then the decision function $f$ will have equal influences for all items. 
\end{proposition}
A proof is in Appendix \ref{proof-sec:equal-degree-influence}. 
A consequence of Proposition \ref{prop:influence-same-degree-same} is that low degree nodes in the graph, i.e., nodes with no or few connections to other nodes, have low influence. This is in line with item-test correlations (or regressions) as described in \citet[][Section 3.7]{Lord:1968}, which would yield coefficients (close to) 0 if the item is not correlated to other items. We give a numerical illustration of this in Section \ref{sec:numerical-illustration}.

%---------------------------------------------------------------------------
\section{Measurement error and test theory}\label{sec:measurement-error-classical}\noindent
The influence gives an indication about the effect a single item can have to change the decision. Here we take this principle a step further and determine stability: the proportion of items necessary to flip that will result in a different decision. To introduce stability and reliability of Boolean functions, we need a process that induces measurement error. Here we introduce such a mechanism. We show that this mechanism has clear connections with test theory. Specifically, the mechanism satisfies the assumptions of classical test theory as described in \citet[][Chapter 2]{Lord:1968}, but the interpretation of the true score is different. 
We also provide a strong and weak version of connections to modern test theory (unidimensional latent variable models), indicating that only for a complete graph do we obtain a unidimensional latent variable representation and that, in this sense, vanishing conditional dependence \citep{Ellis:1997} is a very strong condition. 

%One desirable property of a decision function is that no single item should dominate the decision; otherwise the decision function would not possess the properties of monotony (a) and symmetry (d). In the social theory context the extreme situation is where a decision is determined by a single item, which is referred to as a dictator function. Formally, a dictator function $f$ is such that $f(x)=x_{i}$ for any input $x$ \citep[see, e.g.,][]{Kelly:1988, Garban:2014}. To exclude such extreme cases, we would want that no item has a dominating influence on the decision. 

%----------------
\vspace{1em}\noindent
{\bf Measurement error}\\
We set up a procedure to create a new vector $Y\in \{-1,1\}^n$ (observed variable) from a given original vector $X\in\{-1,1\}^n$, such that independently we flip the random variable (item) $X_{i}$  with probability $\frac{1}{2}(1-\rho)$ with $\rho\in [-1,1]$ \citep{ODonnell:2014,Garban:2014}. We will see that this $\rho$ is proportional to the correlation between $X_i$ and $Y_i$, and hence, the reliability is proportional to $\rho^2$. The obtained data $Y$ from our original variable $X$ represents the measured (contaminated) data, including measurement error. We then consider for the contaminated data $Y$ (observed score) the effect on the decision function $f$, i.e., whether $f(X)=f(Y)$. This allows us to consider how much effect the measurement error has on the decision using $f$. 

To obtain the contaminated data $Y_i$ from the original data $X_i$, we fix the parameter $\rho\in [-1,1]$ for the probability of obtaining the original score $X_{i}$ or its negative $-X_i$. For each $Y_{i}$ for $i=1,\ldots, n$, we iterate the experiment independently and with the same $\rho$, such that
\begin{align}\label{eq:measurement-error}
Y_{i} = 
\begin{cases}
X_{i}	&\text{ with probability } \frac{1}{2}(1+\rho)\\
-X_{i}	&\text{ with probability } \frac{1}{2}(1-\rho)
\end{cases}
\end{align}
So, in expectation, $n\frac{1}{2}(1+\rho)$ of the $Y_{i}$ are the same as $X_{i}$ and $n\frac{1}{2}(1-\rho)$ are the opposite of $X_{i}$. We see that if $\rho=1$, then all items in $Y$ are the same as those in $X$, and so nothing is changed; a value $\rho=-1$ implies that all values will change sign.

%----------------
\vspace{1em}\noindent
{\bf Alignment with classical test theory}\\
To relate the induced measurement error in (\ref{eq:measurement-error}) to classical test theory, we use the true score in the operational view of \citet[][Definition 2.3.1]{Lord:1968}. 
The expectation of the random variable with measurement error $Y_{i}$ conditioned on the value $X_{i}$ is
\begin{align}\label{eq:expectation-y}
\E(Y_{i}\mid X_{i}) = X_{i}\tfrac{1}{2}(1+\rho) + (-X_{i})\tfrac{1}{2}(1-\rho)=\rho X_{i}
\end{align}
where we take the expectation with respect to the measure $\Prob_{\rho}(Y_{i}=X_{i}\mid X_{i})=\frac{1}{2}(1+\rho)$ from (\ref{eq:measurement-error}), which we denote by $\E_{\rho}$, i.e., $\E_{\rho}(X_i):=\E(Y_{i}\mid X_{i})$. We also introduce the measure $\mathbb{P}_{p_i,\rho}:=\mathbb{P}_{p_i}\mathbb{P}_\rho$, and its associated expectation $\mathbb{E}_{p_i,\rho}:=\mathbb{E}_{p_i}\mathbb{E}_{\rho}$. From this, we easily derive the interpretation of the parameter $\rho$ as the  uncentered covariance between $X_i$ and $Y_i$, because
\begin{align*}
\mathbb{E}_{\rho}(X_iY_i) = \tfrac{1}{2}(1+\rho)X_i^2 - \tfrac{1}{2}(1-\rho)X_i^2 = \rho
\end{align*}
since $X_i^2=1$.
%
%%
%\begin{align}\label{eq:true-score-contaminated-score}
%\mu_{\rho}(X_{i})=\E_\rho(Y_{i})=\rho X_{i}=
%\begin{cases}
%\rho		&\text{ if } X_{i}=1\\
%-\rho		&\text{ if } X_{i}=-1
%\end{cases}
%\end{align}
%%
We are then able to show that the assumptions (axioms) to obtain the true score are satisfied by the measurement error defined in  (\ref{eq:measurement-error}).
\begin{lemma}(Alignment with classical test theory)\label{lem:assumptions-classic-tt}
The assumptions of classical test theory, as stated in \citet[][Theorem 2.7.1]{Lord:1968}, are satisfied by random variables $X_i$ with contaminated $Y_i$ and $Z_i$ (independent), using the set-up with measurement error defined by (\ref{eq:measurement-error}), i.e.,  
\begin{itemize}
\item[(i)] the expectation of the error is 0: $\E_{p_i,\rho}(Y_{i}-\mu_{\rho}(X_{i}))=0$,
\item[(ii)] the correlation between the error and the true score is 0: \\$\text{\rm cor}_{p_{i},\rho}(Y_{i}-\mu_{\rho}(x_{i}),\rho X_{i})=0$
\item[(iii)] the correlation between the error and the true score from another measurement is 0: $\text{\rm cor}_{p_{i},\rho}(Y_{i}-\mu_{\rho}(x_{i}),\rho X_{i})=0$ 
\item[(iv)] the correlation between distinct error measurements is 0: \\$\text{cor}_{p_i,\rho}(Y_i-\rho X_i,Z_i-\rho X_i) $
\end{itemize}
\end{lemma}
A proof can be found in Appendix \ref{proof-sec:alignment-ctt}.
%As in \citet{Lord:1968}, by taking the expectation of the observed score, we obtain the true score $\mu_{\rho}(X_{i})$ because the error is 0 in expectation for each value of $X_{i}$ separately. Using a law of large numbers argument, this can be achieved empirically with reasonable accuracy. Also, reliability of parallel tests can be obtained by additionally requiring independence between parallel tests \citep[see also, e.g.,][]{Zijlmans:2018}. Hence, using the setup for measurement error in (\ref{eq:measurement-error}) we obtain the assumptions for the true score as defined in (\ref{eq:expectation-y}) in classical test theory.

Having these assumptions implies that we can determine the reliability between $X$ (the original score) and $Y$ (observed score, contaminated data). The reliability is usually defined as the square of the correlation between $X$ and $Y$ \citep[][Chapter 3]{Lord:1968}. 

\begin{figure}[t]\centering
\begin{tabular}{c @{\hspace{1em}} c}
\pgfimage[width=0.5\textwidth]{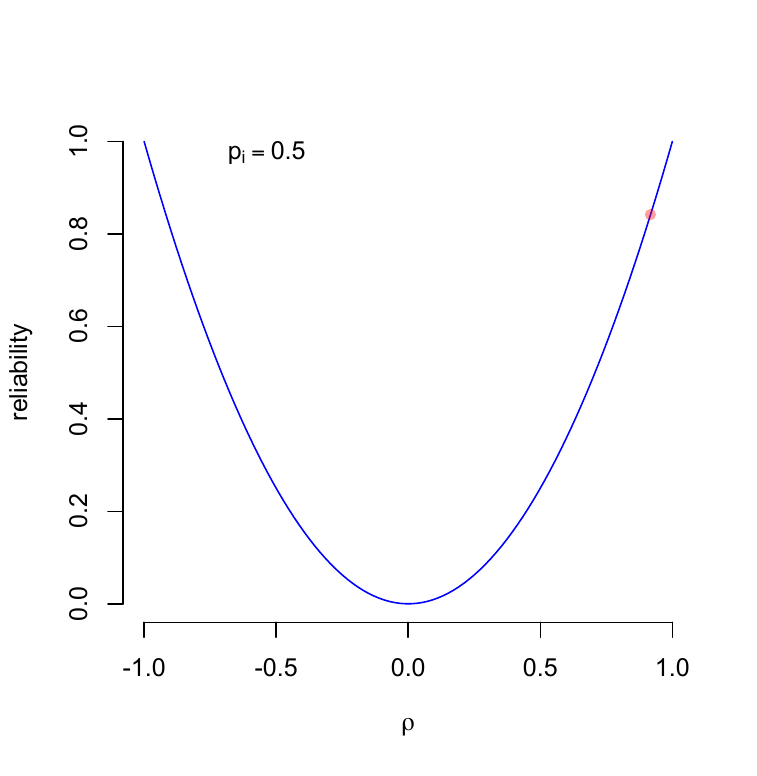}
&
\pgfimage[width=0.5\textwidth]{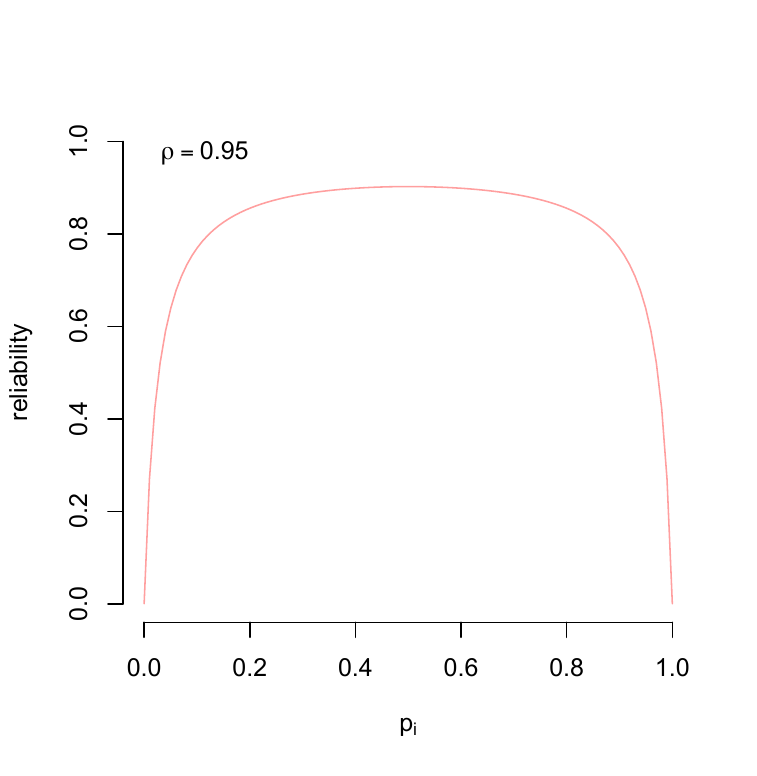}\\[-1em]
(a)	&(b)
\end{tabular}
\caption{Examples of reliability (\ref{eq:reliability}). In (a) reliability as a function of $\rho$, the uncentered covariance between $X_i$  and $Y_i$, for of $p_i=0.50$ in (\ref{eq:p-biased-mean}). The red dot shows the reliability for $\rho=0.95$. In (b) reliability as a function of $p_i$ with $\rho=0.95$, so that $5\%$ is flipped.}
\label{fig:reliability}
\end{figure}

To obtain the correlation between $X_i$ and $Y_i$ we use standard transformations. We have that $\E_{p_i}(X_i)=\mu_i$ and $\E_{p_i}\E_{\rho}(Y_i)=\rho\mu_i$, the variance of $X_i$ is $\sigma_i^2=\text{var}_{p_i,\rho}(X_i)=1-\mu_i^2$, and the variance of $Y_i$ is $\sigma_{i,\rho}^2 = \text{var}_{p_i,\rho}(Y_i)=1-\rho^2\mu_i^2$, with $\mu_i=2p_i-1$ in (\ref{eq:p-biased-mean}). Then we have the transformations
\begin{align}\label{eq:basis-function-y}
\phi(X_{i})=\frac{X_{i}-\mu_{i}}{\sigma_i}
\quad\text{ and }\quad
\phi_{\rho}(Y_{i})=\frac{Y_{i}-\rho\mu_{i}}{\sigma_{i,\rho}}
\end{align}
Since the mean and variance of $\phi(X_i)$ and $\phi_\rho(Y_i)$ are 0 and 1, respectively, we can determine the correlation.
\begin{theorem}\label{thm:reliability} (Reliability)
Given $Y_i$ subjected to measurement error (\ref{eq:measurement-error}) with probability $\tfrac{1}{2}(1-\rho)$ of obtaining $-X_i$, the correlation between $Y_i$ and $X_i$ is 
\begin{align}
%\text{cor}_{p_{i},\rho}(X_{i},Y_{i}) =\E_{p_{i},\rho}\left(\frac{X_{i}-\mu_{i}}{\sigma_{i}}\right)\left(\frac{Y_{i}-\rho\mu_{i}}{\sigma_{i,\rho}}\right)
\text{cor}_{p_{i},\rho}(X_{i},Y_{i}) = \E_{p_{i},\rho}(\phi(X_{i})\phi_{\rho}(Y_{i}))=\rho\frac{\sigma_{i}}{\sigma_{i,\rho}}=\rho\sqrt{\frac{1-\mu_i^2}{1-\rho^2\mu_i^2}}
\end{align}
And the reliability is
\begin{align}\label{eq:reliability}
\text{cor}_{p_i,\rho}^2(X_i,Y_i)=\rho^2\frac{\sigma_{i}^2}{\sigma_{i,\rho}^2}=\rho^2\frac{1-\mu_i^2}{1-\rho^2\mu_i^2}
\end{align}
\end{theorem}
A proof is in Appendix \ref{proof-sec:reliability}. We observe that the reliability in (\ref{eq:reliability}) is similar to the one given in \citet[][Chapter 3]{Lord:1968}, the ratio of the true score variance to the observed score variance, except that here reliability is multiplied by $\rho^2$, the squared uncentered covariance between original score $X_i$ and contaminated score $Y_i$. Additionally, this squared uncentered covariance $\rho^2$ attenuates $\mu_i^2$ in the denominator. An example of reliability in (\ref{eq:reliability}) as a function of $\rho$ is shown in Figure \ref{fig:reliability}(a) for $\mu_i=2p_i-1$ given in (\ref{eq:p-biased-mean}) with $p_i=0.5$. The red dot shows the reliability for $\rho=0.95$, corresponding to the maximal value in (b) at $p_i=0.5$. In Figure \ref{fig:reliability}(b) is the reliability as a function of $p_i$ for $\rho=0.95$.

\vspace{1em}\noindent
{\bf Differences with classical test theory}\\
There are several consequences of imposing this measurement error mechanism on the original score to obtain observed data. 

\textit{Effect of $\rho$}. For binary data it becomes clear how the measurement error affects the reliability; there is attrition in the reliability, as expected, from the mechanism to introduce measurement error. Clearly, if there is no measurement error, then $\rho=1$, and reliability is 1. Additionally, $\rho$ enters the denominator of (\ref{eq:reliability}). Since $\rho\in [-1,1]$, this leads to an increase in reliability. And so the factors counteract, as can be seen in Figure \ref{fig:reliability}.

\textit{True scores}. 
%The mechanism to introduce measurement error is no longer restricted to using the additive structure of the true score and error at the level of the sum score only \citep{Lord:1968}. In other words, a mechanism to introduce measurement error is not restricted to linear (continuous) measurement. 
There are differences in interpretation of the true score between classical test theory and the theory presented here. \citet[][Chapter 2]{Lord:1968} define the true score as the expectation of the observed score $Y_i$, i.e., $\mathbb{E}(Y_i)=\tau$, while in our framework, the original score $X_i$ is the true score, and hence, in the present version is either $-1$ or $1$. This has the following implications for the expectations.   \\
\begin{itemize}

\item[($a$)] \textit{True score from linear model} Let $Y_i$ be the observed value, where $Y_i$ is obtained from the linear model $X_i+e_i$, where the random variable $e_i$ is independent of $X_i$, has expectation 0 and finite variance. Then the true score of observation $Y_i$ is $\E(Y_{i})=\tau$ \citep[][Definition 2.3.1]{Lord:1968}.

\item[($b$)] \textit{True score from probabilistic model} Let $Y_i$ be the observed value, where $Y_i$ equals $X_i$ with probability $\tfrac{1}{2}(1+\rho)$ and $\rho\in [-1,1]$, as defined in (\ref{eq:measurement-error}). Then the true score of observation $Y_i$ is $X_i$, and the expectation is $\E_{p_i,\rho}(Y_{i})=\rho \mu_i$.\\

\end{itemize}
We observe that the difference in true score is that in definition ($a$), that of \citet{Lord:1968}, only exists as the expectation (a continuous value), while in our framework, the true score is simply the original value $X_i$, either $-1$ or $1$. Subsequently, we can connect the expectations of the two frameworks in ($a$) and ($b$) by stating that 
\begin{align}
\mathbb{E}_{p_i,\rho}(Y_i)=\mathbb{E}_{p_i}\mathbb{E}(Y_i\mid X_i)=\rho\tau
\end{align}
where $\tau$ is interpreted as $\tau=\mathbb{E}(X_i)$, which in our framework is the conditional expectation $\mu_i=\mathbb{E}(X_i\mid X_{\partial i})$ given in (\ref{eq:p-biased-mean}), using the measure in (\ref{eq:ising-conditional-probability}). Clearly, only if $\rho=1$, then the two definitions align.

%Interestingly, the true score according to the probabilistic model, $\rho X_i$, is in the interval $[-1,1]$. When $\rho=1$, and $\rho X_i$ is $-1$ or $1$, then there is no measurement error and we obtain the same as the true score from the additive model in ($a$). When $\rho = -1$, then we can be certain that the value $X_i$ has been flipped. When $\rho$ is near 0 (negative or positive), then the original value $X_i$ is almost as likely to be negative or positive; there is much uncertainty about the sign of $X_i$. The true score $\rho X_i$ can therefore be interpreted as an indication of certainty of the measurement error. Our main focus here is, however, to consider what happens to the decision $f(Y)$, when using the observed (contaminated) data, with respect to the decision $f(X)$, when using the original (true) data. 
%This implies that true score in the probabilistic from ($b$) can also be considered as an additional process on top of that of ($a$). 

\textit{Konwledge space}. Our approach is also different from one based on knowledge space theory, where the size of a knowledge structure determines the true score of an individual \citep{Chiusole:2024}. The traditional reliability indices will then possibly be larger than one, and so information theoretic indices were introduced, that  were shown to be more consistent with underlying reliability of knowledge space theory.  

\textit{Effect on decision}. A final difference of imposing this mechanism for measurement error, is that it can be used to determine the stability of a decision based on the observed data (together with the Fourier analysis of Boolean functions). This last point will also lead to a more thorough and more general investigation into the motivation for using the sum score in applied research (see Section \ref{sec:merit-sum-score}).
 
 %-----------------
 \vspace{1em}\noindent
 {\bf Alignment with modern test theory}\\
A relation to modern test theory can be found through the properties of conditional association and vanishing conditional dependence. A vector $X$ that satisfies these two properties has been shown to be equivalent to a unidimensional and monotone latent variable representation, such that all $X_i$ are conditionally independent given the latent variable \citep[i.e., local independence][]{Ellis:1997}. Here we combine this result with that of \citet{Castillo:1998}, which shows that the marginal over the graph of this latent variable must be a fully connected graph. This connects to the equivalence between the Ising model and the latent variable representation, described in \citet{Marsman:2018}.

Intuitively, conditional association says that splitting up a graph (or vector of nodes) in two parts, implies that there has to be a positive connection between the two parts for a positive correlation.  \citet[][Theorem 6]{Holland:1986b} state that, if we assume that a latent variable representation holds, such that it has local independence, is unidimensional, and monotone (referred to as the three properties), then conditional association is implied. The three properties are sufficient for conditional association; in other words, if you do not have conditional association, then you do not have a latent variable representation which has the three properties. More is needed when going in the other direction. This is given by vanishing conditional dependence \citep{Ellis:1997}. Vanishing conditional dependence is a stronger statement, implying that there exist infinitely many $X_i$ (so the graph $G$ has infinite size) such that there exists a latent variable representation with the three properties.  In other words, any part of the graph outside the nodes of interest has to be fully connected to the graph of interest to remove all correlations between the nodes. 

More formally, we have the following definitions. Conditional association is defined as follows: For any partition $K$ and $L$ of the nodes $V$ of the graph $G$, we obtain that $\text{cov}_{\pi,\rho}(f(X_K),g(X_K)\mid h(x_L))\ge 0$ for monotone and bounded functions $f$ and $g$, conditional on a discrete outcome function $h:\{-1,1\}^{k}\to D\subset \mathbb{Z}$, such that $h(x_{L})=c$  \citep{Holland:1986b,Junker:1993}. A vector $X_V$ has vanishing conditional dependence if for any partition $K$, $L$ and $W$ of $V$, with $|V|\to\infty$, it holds for any measurable $f,g$ that $\text{cov}_{\pi,\rho}(f(X_K),g(X_L)\mid X_W)\to 0$ as $|W|\to \infty$ \citet{Ellis:1997}. Then the following corollary follows from Theorem 5 in \cite[][]{Ellis:1997}
and Theorem 1 in \citet[][]{Castillo:1998}.

\begin{corollary} \label{cor:vanishing-conidtional-dependence}
Let $X$ be the random variables induced by the Ising probability in (\ref{eq:ising-joint-probability}) with respect to $G=(V,E)$ with $|V|\to \infty$ and positive connectivity parameters. Then $X$ is conditionally associated and has the property of vanishing conditional dependence if and only if the graph $G$ is fully connected. 
\end{corollary}
A proof is in Appendix \ref{proof-sec:vanishing-conidtional-dependence}. The work by \citet{Marsman:2018} shows the equivalence of the fully connected Ising model to the marginal Rasch model, a well-known model in item response theory \citep[e.g.,][]{Linden:2013}. Conceptually, this tells us that the coherence between the items must be so strong, that it could be represented by a single latent variable connected to all nodes of the graph; this in turn implies that the marginal over the graph is then fully connected. 

\vspace{1em}\noindent
{\bf Differences with modern test theory}\\
The result in Corollary \ref{cor:vanishing-conidtional-dependence} is rather limited in that the graph must be fully connected; missing a single connection already breaks the corollary, so that there is no latent variable representation. Satisfying only the conditional association property leads to much weaker constraints on the graph, but, obviously, also leads to the weaker statement of conditional association. Essentially, to obtain conditional association, the graph needs to be connected, i.e., you must be able to get from any node to any other node, and the coefficients of an LTF must be well chosen.

\begin{theorem}\label{thm:conditional-association}
Let $G$ be a graph with nodes $X$ that follows the Ising probability distribution (\ref{eq:ising-joint-probability}) with equal external field $\xi$ and positive connectivity parameters $\theta_{ij}$ for all $i,j\in V$. Furthermore, let $h:\{-1,1\}^{k}\to D$, with $D\subset \mathbb{Z}$ be some discrete outcome function Then: 
\begin{itemize}
\item[(i)] If $G$ is disconnected (has two or more components), then $X$ defined on $G$ is not conditionally associative. 
\item[(ii)] If $G$ is connected, and the Boolean functions $f$ and $g$ are LTFs with coefficients $(a_i,i\in V)$ and $(b_i,i\in V)$, respectively, such that $\sum_{i\in K} a_i b_i\sigma_i^2\ge 0$ for any finite subset $K$ of $V$, then $X$ is conditionally associative. 
%\item[(3)] If $G$ is fully connected with positive interaction parameter, then $X$ defined on $G$ is conditionally associative. 
\end{itemize}
\end{theorem}
A proof can be found in Appendix \ref{proof-sec:conidtional-association}. This weaker condition allows you to rule out conditional association by knowing that there are two components in the graph. Theorem \ref{thm:conditional-association} shows that the interplay between the structure of the graph and the conditions for an item response model make this approach interesting.

%---------------------------------------------------------------------------
\section{Reliability and stability of decisions}\label{sec:stability-reliability}\noindent
Now that we set up the mechanism to obtain data $Y_{i}$ with measurement error defined in (\ref{eq:measurement-error}), we move on to its effect on decisions. We are interested in what the result is on the decision of measurement error with respect to the original data defined in (\ref{eq:measurement-error}), i.e., will the decision remain the same upon changing a proportion $\tfrac{1}{2}(1-\rho)$ of the original score $X_{i}$. 

\vspace{1em} \noindent
{\bf Reliability}\\
Applying the Fourier expansion in (\ref{eq:p-biased-fourier-coefficient}), we can determine the reliability (squared correlation) of the decision function $f$ with input $X$ and decision function $f$ with input $Y$, using the correlation $\text{cor}_{\pi,\rho}(f(X),f(Y))$. 

%---------------------------------------------------------------------------
\begin{theorem}\label{lem:variance-covariance-f(x)-f(y)}(variance and covariance of $f(Y)$ and $f(X)$)
Let $f:\{-1,1\}^{n}\to \{-1,1\}$ be a Boolean function and for all $i$, $Y_{i}$ are obtained from the experiment where with probability $\tfrac{1}{2}(1+\rho)$, $Y_{i}=X_{i}$ and with probability $\tfrac{1}{2}(1-\rho)$, $Y_{i} =-X_{i}$ for each $i$ independently. Then the variance of $f(Y)$ is 
\begin{align*}
\text{var}_{\pi,\rho}(f(Y)) = \sum_{S\ne \varnothing} \hat{f}^{\pi,\rho}(S)^{2}
\end{align*}
and the covariance between $f(X)$ and $f(Y)$ is 
\begin{align}\label{eq:covariance-measurement-error}
\text{cov}_{\pi,\rho}(f(X),f(Y)) = \sum_{S\ne \varnothing} \omega(S)\rho^{|S|}\hat{f}^{\pi}(S)\hat{f}^{\pi,\rho}(S)
\end{align}
where $\omega(S)=\prod_{i\in S}\tfrac{\sigma_{i}}{\sigma_{i,\rho}}$.
\end{theorem}
A proof is in Appendix \ref{proof-sec:var-cov-f}.
We see that if $\rho$ is not too large the reliability of the decision function is mostly determined by subsets of order $|S|\le 1$, because of the factor $\rho^{|S|}$ in (\ref{eq:covariance-measurement-error}). 
However, if we additionally assume that $f$ is an LTF, we immediately obtain that only the coefficients for $|S|\le 1$ are required \citep[][Theroem 5.1, Chow's theorem]{ODonnell:2014}. We can therefore use an approximation with only the singleton sets $S=\{i\}$ to use for the correlation.

\begin{proposition} (Approximate decision reliability) \label{prop:approx-decision-reliability}
Let $f$ be an LTF Boolean function and $Y$ be an $n$ vector defined in (\ref{eq:measurement-error}) such that the correlation is $\text{cor}_{p_{i}\rho}(X_{i},Y_{i})=\omega(i)\rho$ for all $i$. Then the decision reliability is the square of
\begin{align}\label{eq:reliability-monotone-functions}
\text{cor}_{\pi,\rho}(f(X),f(Y)) = \rho\sum_{i\in V} \omega(i)\hat{f}^{\pi}(i)\hat{f}^{\pi,\rho}(i)+ O(\rho^{2}/2)
\end{align}
\end{proposition}
See Appendix \ref{proof-sec:approx-dec} for a proof. 
This approximation is for general Boolean functions very accurate whenever $\rho$ is relatively small, since then the remainder term $O(\rho^{2})$ is small. However, for LTF functions, such as the majority function, the Fourier coefficients on the singleton sets, i.e., $f(i):=f(\{i\})$, are relatively high compared to larger sets with $|S|>1$ \citep[][Theorem 5.2]{ODonnell:2014}. This is also evidenced by the bound we obtained in (\ref{eq:bound-influence}). Hence, for functions satisfying monotonicity (a) and transitive symmetry (e), this approximation is quite accurate. 

We see from (\ref{eq:reliability-monotone-functions}) that the reliability of decision function $f(Y)$, that is $\text{cor}_{\pi,\rho}^2(f(X),f(Y))$, is proportional to $\rho^2$, with $\rho$ the parameter that induced the measurement error, and the reliabilities of the items in $\omega^2(i)$. This implies that the reliabilities of the items induce the decision reliabilities, distorted by the measurement error with parameter $\rho$ and the properties of the decision function in the Fourier coefficients $\hat{f}^{\pi}(i)$ and $\hat{f}^{\pi,\rho}(i)$. 

Using (\ref{eq:reliability-monotone-functions}) we can determine the reliability of our decision relatively easily in practice.
This approximation is useful since we only require estimation of singleton Fourier coefficients (we use this fact in the numerical illustrations in Section \ref{sec:numerical-illustration}).

\vspace{1em} \noindent
{\bf Stability}\\
The correlation between $f(X)$ and $f(Y)$, where $Y$ is the contaminated input of $X$, given in (\ref{eq:reliability-monotone-functions}) can be interpreted in another way, which we find useful with respect to the properties (a)-(e). The impact of measurement error on the decision can also be defined by comparing the probabilities of $f(X)=f(Y)$ and $f(X)\ne f(Y)$ \citep{ODonnell:2014}. This difference in probabilities is called stability \citep{ODonnell:2014} or noise stability \citep{Garban:2014}.

More formally, let $Y$ be a contaminated version of the original $X$ as defined in (\ref{eq:measurement-error}) and let $f$ be a Boolean function. Then the  {\em stability of $f$ at $\rho$} is 
\begin{align}\label{eq:stability}
\mathbb{S}_{\pi,\rho}(f) = \Prob_{\pi,\rho}(f(X)=f(Y)) - \Prob_{\pi,\rho}(f(X)\ne f(Y))
\end{align}
A more precise statement of (f) in Section \ref{sec:boolean-functions} is then that a decision function $f$ is \\
\begin{itemize}
\item[(f)] {\em stable} if $Y$ equals $X$ with probability $\tfrac{1}{2}(1+\rho)$ for each coordinate such that the probability that $f(X)=f(Y)$ is at least $1-\epsilon$, for small $\epsilon>0$, i.e., when the stability is $\mathbb{S}_{\pi,\rho}(f)\ge 1-\epsilon$. \\
\end{itemize}
%
%stable if the probability of $f(X)=f(Y)$ is at least $1-\epsilon$, for some $\epsilon>0$, i.e., when the stability is $\mathbb{S}_{\pi,\rho}(f)\ge 1-\epsilon$.
%And so the probability that decisions are equal according to $f$ is 
%%
%\begin{align*}
%\Prob_{\pi,\rho}(f(X)=f(Y))= \frac{1}{2}+\frac{1}{2}\mathbb{S}^{\pi}_{\rho}(f)
%\end{align*}
%%
This implies that the probability of a mismatch between the decision based on the original score $X$ and the contaminated score $Y$ should be small. 

There is a direct relation between stability and reliability. Because $f(X)f(Y)$ is 1 whenever $f(X)=f(Y)$ and $-1$ whenever $f(X)\ne f(Y)$, we can rewrite the stability in (\ref{eq:stability}) using the expectation operator. We then obtain that $\mathbb{S}_{\pi,\rho}(f) = \E_{\pi,\rho}(f(X)f(Y))$. This implies that
\begin{align}\label{eq:stability-reliability}
\mathbb{S}_{\pi,\rho}(f) = \text{cor}_{\pi,\rho}(f(X),f(Y))
\end{align}
for mean zero decision functions. In other words, the stability is the same as decision function correlation for unbiased decisions, which implies that reliability of decision functions is the square of stability. Hence, property (f) about stability, is in fact also about the reliability between the decisions based on $X$ and $Y$. Because we believe that reliability is an important property, it follows that stability is also an important property. We prefer to refer to stability because it relates more closely to the modern concept of a noise induced process and its relation to the true underlying process, which is what we are trying to get at.

%---------------------------------------------------------------------------
\section{Evaluation of properties (a)-(f)}\label{sec:merit-sum-score}\noindent
We now have a framework that allows us to evaluate the properties for Boolean functions that serve as decision functions. Here, we will investigate what type of decision function $f$ is optimal with respect to several desirable (minimal) properties one could wish for decision functions. 
The properties we deemed important in Section \ref{sec:boolean-functions} are (a)-(e) described in Section \ref{sec:boolean-functions}, which we would like to augment with (f) stability. We argue for this in this section and simultaneously obtain, as a corollary, that the (weighted) sum score with threshold is actually the most favourable (stable) decision function.

In this section we will elaborate on two types of arguments that seem convincing to us. These two arguments focus on property (f) stability of the decision function and on a weaker version of property (c), unanimity of the decision function. But, in using monotonicity (a) and permutation symmetry (e), leading to the convenient characterisation of stability (the square root of decision function reliability) in (\ref{eq:reliability-monotone-functions}), we already have properties (a), (c), and (e). Then by May's theorem \citep[][and Appendix \ref{proof-sec:may-thm}]{May:1952}, (b) and (d) are unique to simultaneously satisfy in an LTF (a weighted sum score with decision). Hence, all we need to show is that stability (f) is a useful property. 
Both our arguments apply to the situation where the influence of each item on the decision $f$ is upper bounded, which seems reasonable given Proposition \ref{prop:bound-influence}. Our two arguments are as follows:
\begin{itemize}
\item[(1)] an LTF is insensitive to noise, i.e., is stable (property (f));
\item[(2)] a decision based on an LTF matches the input best (property (c));
%\item[(3)] a weighted sum score aligns with classical and modern test theory.\\
\end{itemize}
We shall discuss each argument in turn.

%-----------------------------------------------------
\vspace{1em}
\noindent
{\bf The (weighted) sum score decision function is stable}\label{sec:ltf-stable}\\\noindent
Stability was defined in (\ref{eq:stability}) as the difference between the probability of $f(Y)=f(X)$ and $f(Y)\ne f(X)$, where $X$ is the original score and $Y$ the observed score.  We argue that stability (directly related to reliability at the level of the decision in (\ref{eq:stability-reliability})), in terms of changing the original score with some small percentage, is an important property and the sum score is the decision function that is most stable. We believe this is reasonable because, if the measurement error, as induced by (\ref{eq:measurement-error}), is small (with $\rho$ close to 1), then we expect the decision based on the observed score $Y$ to be the same as the decision based on the original score $X$. And this is exactly what stability (reliability) at the level of the decision tells us. 

The stability of the decision $f(Y)$ with respect to the decision $f(X)$ can be shown to be best for an LTF in the case where the influence of each item is upper bounded. We obtained such a bound in Proposition \ref{prop:bound-influence}. Then, it can be shown that  $\mathbb{S}_{\pi,\rho}(\text{maj}_{n})\ge \mathbb{S}_{\pi,\rho}(f)$ for any $f$, where $\text{maj}_n=\text{sign}(X_1+\cdots + X_n)\in \{-1,+1\}$ is the majority function, an LTF used for a decision \citep{Mossel:2010}. 
In particular for standardised variables, such that the reliability is $\rho^2$, the stability of the majority function $\text{maj}_{n}$ is \citep[][Theorem 5.17]{ODonnell:2014}
\begin{align}\label{eq:stability-majority}
\mathbb{S}_{\pi,\rho}(\text{maj}_{n}) = \frac{2}{\pi}\arcsin \rho \quad \text{as } n \to \infty
\end{align}
Stability $\mathbb{S}_{\pi,\rho}(\text{maj}_n)$ for large $n$ takes values in $[-1,1]$ (because of the scaling $2/\pi$). 
This implies that for any function $f$, changing a proportion $\tfrac{1}{2}(1-\rho)$ of the original scores results in the largest stability for the LTF compared to any other decision function. 
%The assumptions in \citet{Mossel:2010} to obtain (\ref{eq:stability-majority}) are that the influence of each item is approximately similar, and that the expectation of the Boolean function (here $\text{maj}_{n}$) is 0. Then it is shown that for any other function than the majority function, the stability of the majority function is higher. The implication is that, with respect to measurement error as defined in (\ref{eq:measurement-error}), a weighted sum score of the items (which have approximately the same influence) is the stablest function among all functions. This means that using a weighted sum score as a decision function, will, when a small percentage of the items have been flipped (measurement error), not immediately  lead to a different decision; and there is no other function that will be better in this respect. 
So, in this sense, the stability, and hence the reliability, at the decision level is highest for an LTF compared to any other decision function. 
\vspace{1em}\noindent
{\bf The decision of a (weighted) sum score matches the input best}\label{sec:rousseau}\\ \noindent
Property (c) in Section \ref{sec:boolean-functions} is about unanimity, it is the extreme case where the decision is $-1$ if all input is $-1$, or the decision is $1$ if all input is $1$. We are looking for a less extreme version of this property. If one of the coordinates $X_i$ were flipped, then we would in most cases still want the decision to be the same, unless we purposefully give an item a large influence. If most of the items $X_i$ are $1$ then the decision should be $1$; the input and the decision should agree. This property is sometimes referred to as Rousseau's property  \citep[see, e.g.,][]{Schwartzberg:2008} or Rousseau's Theorem  \citep[][Theorem 2.33]{ODonnell:2014}.

According to Rousseau's Theorem, the agreement between the decision and the input (items) is maximised by an LTF. In the case of uniform probability over $\{-1,1\}$, the expected number of items to agree with the decision $f(X)$ is $\frac{n}{2}+\frac{1}{2}\mathbb{I}_{\frac{1}{2}}(f)$ for monotone functions \citep[see][Proposition 2.32]{ODonnell:2014}, indicating that one would expect at least half of the items to point in the same direction as the decision, which is a reduced version of the unanimity property (c). From this expected number of items that agree with the decision, we see that we can interpret Rousseau's property as having at least half the items in line with the decision and more items will agree with the decision corresponding to the total influence of the items on the decision. 

\section{Numerical illustrations}\label{sec:numerical-illustration}\noindent
To show the usefulness and accuracy of the concepts of Fourier analysis and stability, we consider some numerical illustrations. We show with simulations that even though the graph structure is not accurately recovered, the Fourier coefficients are still accurate. 

In Appendix \ref{sec:algorithm-fourier-analysis} we describe the algorithm to obtain the estimates of Fourier coefficients, using penalised pseudo-likelihood. We also show that the convergence to the true values of the Fourier coefficients is at the rate of $1/\sqrt{m}$, where $m$ is the number of replications. Here we illustrate several characteristics of the framework with numerical illustrations. 

\begin{figure}[t]\centering
\begin{tabular}{c @{\hspace{3em}} c}
	\includegraphics[width=0.45\textwidth]{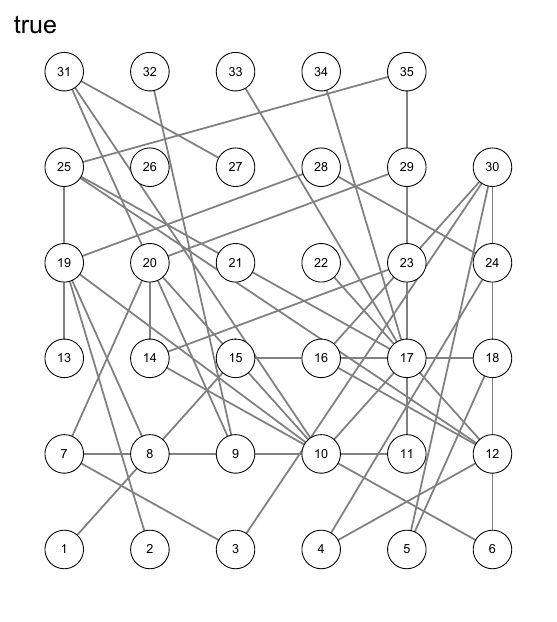} &
	\includegraphics[width=0.5\textwidth]{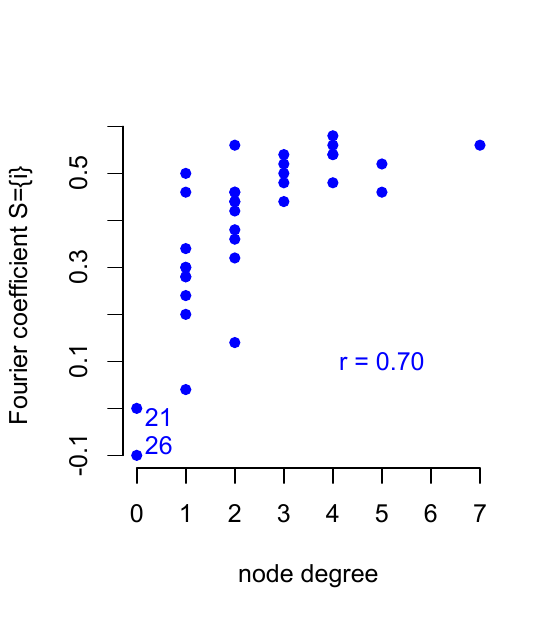}\\
	(a) & (b)
\end{tabular}
\caption{Visualisation of the example random graph in (a) and the scatterplot for the degree of the nodes of the graph in (a) and the singleton Fourier coefficients in (b). The correlation is 0.70. Nodes 21 and 26 have the lowest influence and can be seen to be isolated in the network, literally having no influence on the network. }
\label{fig:example-graph-correlation-fc}
\end{figure}

We start with a single case to illustrate how a Fourier analysis connects to a graphical model and stability. Here we generate binary data according to an Ising model with $n=35$ items and $m=100$ observations according to a random graph with probability of an edge 0.05. Each nonzero edge has weight $\theta_{ij}=3$ and the thresholds are $-\frac{1}{2}\sum_{i=1}^{n}\theta_{ij}$ for each $j=1,2,\ldots, 35$. The data are generated with {\em IsingSampler} \citep[described in][]{Borkulo:2014} in R \citep{R:2012}. The parameters of the conditional Ising model were estimated by {\em IsingFit} \citep{Borkulo:2014}, which estimates the parameters in a nodewise fashion. We used $\gamma=0.25$ for the extended BIC model selection to obtain the appropriate lasso penalty (only on edge parameters). An example of the graph that generated current data is shown in Figure \ref{fig:example-graph-correlation-fc}(a). Estimates of the graph parameters are accurate when signal to noise ratios are high as they are here (see Monte Carlo simulations below). 
%

%To compute the Fourier coefficients we determined for each $i\in V$ the transformed score $\phi(y_{i})$, where $y_{i}=2x_{i}-1$ to transform them from the domain $\{0,1\}$ to $\{-1,1\}$. The conditional Ising probabilities are adjusted accordingly (plugging in the transformed values $x_{i}=\frac{1}{2}(y_{i}+1)$). 
The decisions for each observation $1,2, \ldots, m$ were determined with a LTF with $a_{0}=-0.6$ (pass at 60\% correct) and $a_{i}=\frac{1}{n}$ (equally weighted) for all $i$. The Fourier coefficients were then determined using (\ref{eq:p-biased-fourier-coefficient}) only for singleton sets $S=\{i\}$. The Fourier coefficients can be seen in the scatterplot in Figure \ref{fig:example-graph-correlation-fc}(b) as a function of node degree. It is clear that the coefficients correlate highly with the degree of the nodes in the graph, confirming Proposition \ref{prop:influence-same-degree-same}. The correlation between the Fourier coefficients and degree of this particular graph is 0.70, as expected from Proposition \ref{prop:influence-same-degree-same}. It can also be seen that the isolated nodes, nodes 21 and 26, have the lowest influence (which equals the Fourier coefficient for singleton sets) of 0.0 and -0.1, respectively. This is exactly what can be expected from Proposition \ref{prop:influence-same-degree-same}, since the values of the isolated nodes are independent from the rest, and so will have very little influence on the decision based on the nodes that are connected to each other. 

\begin{figure}[t]\centering
\begin{tabular}{c @{\hspace{3em}} c}
	\includegraphics[width=0.45\textwidth]{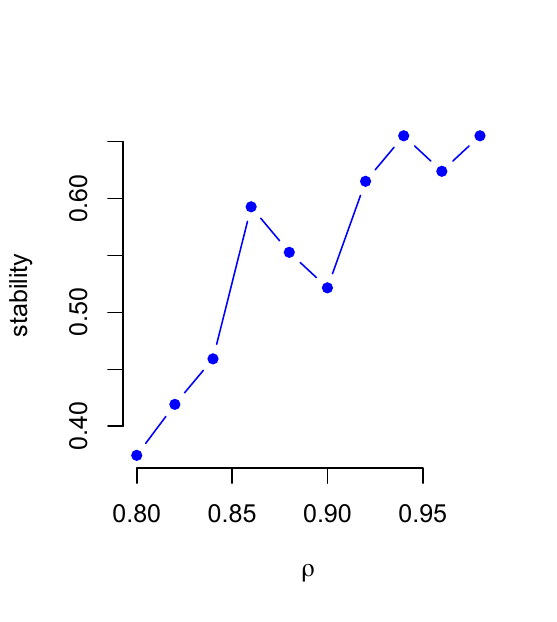}
	&
	\includegraphics[width=0.45\textwidth]{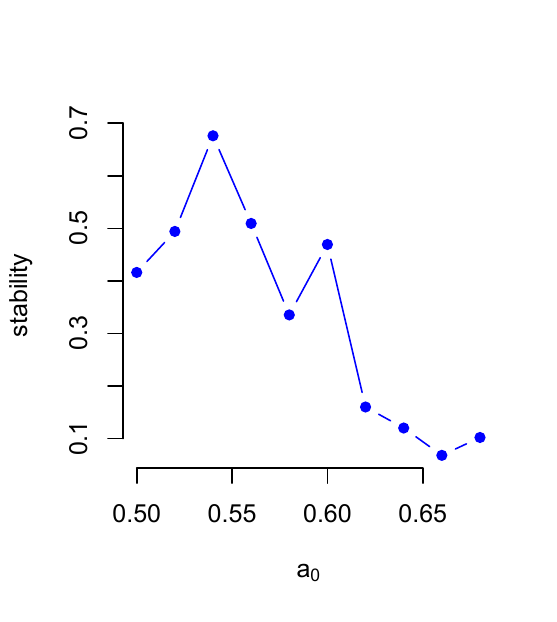}\\
	(a) & (b)
\end{tabular}
\caption{Stability of the small example data as a function of the correlation $\rho$ between the original and corrupted data (a), and as a function of the threshold $a_{0}$ (b). }
\label{fig:example-stability}
\end{figure}

Turning to stability, we are considering whether the rule we chose (an LTF with $a_{0}=-0.6$, so that 60\% is required to be 1) is stable with respect to measurement error. We apply (\ref{eq:measurement-error}) to the generated data and then compute the stability using the Fourier coefficients in (\ref{eq:p-biased-fourier-coefficient}). Figure \ref{fig:example-stability}(a) shows the stability as a function of the measurement error parameter $\rho$ (see (\ref{eq:measurement-error})). It clearly shows, as expected, that the stability increases as the parameter $\rho$ increases. We also considered the threshold $a_{0}$ for several values, shown in Figure \ref{fig:example-stability}(b). This indicates that stability decreases with increasing threshold. An explanation is that at high thresholds the values in $Y$ never reach it and so the decisions $f(Y)$ are nearly constant, resulting in near 0 correlation. 

\begin{figure}[t]\centering
\begin{tabular}{c @{\hspace{3em}} c}
	\includegraphics[width=0.45\textwidth]{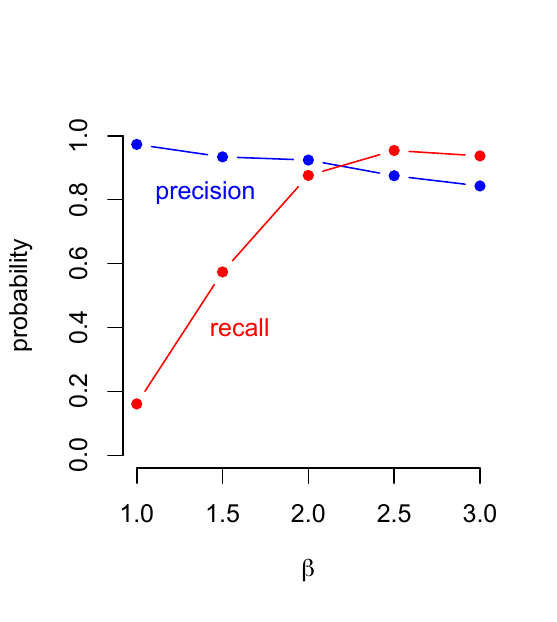}
	&
	\includegraphics[width=0.45\textwidth]{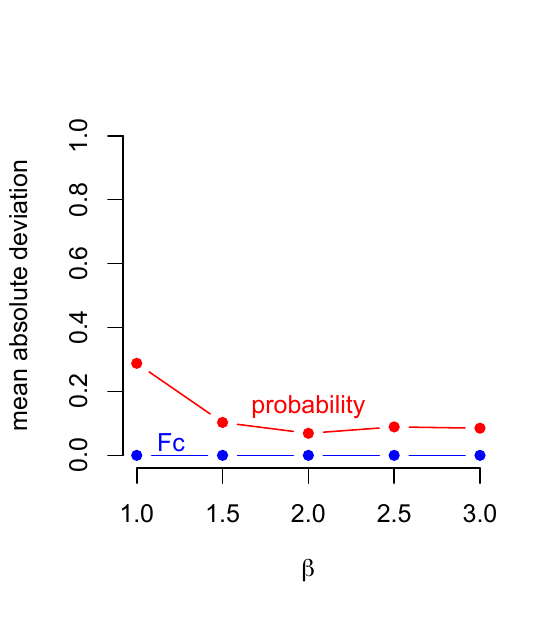}\\
	(a) & (b)
\end{tabular}
\caption{Recovery of edge parameters in terms of precision (blue) and recall (red) as a function of the edge weights ($\beta$) shown in (a). In (b) is the mean absolute deviation of the conditional probabilities (red) and Fourier coefficients (Fc, blue).  }
\label{fig:example-monte-carlo}
\end{figure}

To study the accuracy of the estimates of the Fourier coefficients we use Monte Carlo simulations. The Fourier coefficients depend on the estimates $\hat{p}_{i}$ of the conditional probabilities that are used for the $p$-biased parity functions $\phi$. For all simulations we used $n=35$ items and $m=100$ observations and we varied the signal to noise ratio by increasing the edge weights ($\beta$) from 1 to 3. To determine the accuracy of the graph we determined whether we obtained neighbourhoods accurately and considered the adjacency matrices of the estimated graphs. As dependent measures we used 
\begin{align*}
\text{precision} = \frac{\#(\text{correctly identified edges})}{\#(\text{identified edges})}
\end{align*}
 and 
 \begin{align*}
 \text{recall} =  \frac{\#(\text{correctly identified edges})}{\#(\text{correct edges})}
 \end{align*}
  We see in Figure \ref{fig:example-monte-carlo}(a) that the edge parameters are accurately estimated when the edge weights are large ($\beta\ge 2$), but are estimated poorly when the edge weights are lower. Although precision remains high, the number of correctly recovered edges (recall) is low for low signal to noise ratios. However, in Figure \ref{fig:example-monte-carlo}(b) we see that the mean absolute deviations of the Fourier coefficients remain small, regardless of the effect size $\beta$. This is because the conditional probabilities are still reasonable at low signal to noise ratios. So, even though the graph itself is not accurately recovered, the predictions for the probabilities are reasonable and hence the Fourier coefficients are accurately estimated. The fact that the graph need not be correctly recovered for the conditional probabilities is that the edge weights and threshold parameters are exchangeable in the Ising model. In \citet{Waldorp:2019} we explain in more detail the relation between prediction and graph recovery. 
%

%---------------------------------------------------------------------------
\section{Conclusions and discussion}\noindent
We introduced a new framework for measurement error and showed that this definition aligns with the classical concept of measurement error. A difference between this new framework and classical test theory lies in the interpretation of the true score. This framework allowed us to derive a property of decision functions based on items: stability. This property says that when a small fraction of the items is changed, then the decision should remain the same; otherwise the decision function is sensitive to measurement error. We then used this property to argue for the claim that a (weighted) sum score is the best decision function because (1) it is the most stable decision function, and (2) a (weighted) sum score matches the input best. 

The mechanism for measurement error has a parameter $\rho$ that determined the probability of changing the original score. The parameter $\rho$ surfaced in the derivation of reliability, making clear how measurement error leads to attrition of reliability directly. Currently, we are agnostic as to how $\rho$ comes about, but knowledge of any such process can be built in the theory.  

%Using Fourier analysis of Boolean functions allowed for the investigation of decision reliability. Decision reliability was shown to be induced by item reliability and is affected by both the measurement error and the properties of the decision function, expressed in terms of the Fourier coefficients of the decision function. Based on these analyses, we argued that a weighted sum score satisfies the properties (a)-(f), which is likely to be the only function that satisfies these properties (see also May's theorem), and that therefore the (weighted) sum score is a reasonable way to aggregate items to make decisions. 

The new framework used a graphical model (here the Ising model) to capture dependency and bias (non-zero mean). This connection provided a representation of a test in terms of a graph. We then showed that the Fourier coefficients can be interpreted as the influence of an item on the decision based on the test, which is directly related to an edge to other nodes in the test. This makes it possible to determine if there are any items that seem inappropriate. We showed that small influence (relative to the other items) is an indication of an `isolated' item in a graph; an item that has few or no connections to other items in the graph and so should not be considered as covering same topic. This suggests a picture of a test as a graph of items and their connections, where the degree of the nodes are related the influence of the item by Proposition \ref{prop:influence-same-degree-same}.

Ultimately, our interest was to determine whether a decomposition of decision (Boolean) functions would lead to insights into the properties decision functions should have. We believe our framework indeed provides such insights by suggesting that stability of decisions is an important property. 

We hope that this new framework leads to fruitful research and new concepts within the world of test theory. For example, it is possible within this framework to consider different properties than those discussed here, that lead to appropriate decisions. It may turn out that certain properties are appropriate depending on the context in which they are used, and the context could also be represented in the graph, showing the connections to items in the test and its effect on the decision. 

Another direction is extending the current version for two-category items to multi-category or continuous items. Such an extension would make the presented framework more widely applicable. The results on the relation between unidimensional latent variable models and graphs also leads to the question whether multi-dimensional constructs could be represented and analysed using our framework. This extension is most interesting and may reveal new insights into how constructs might be related and investigated from a different perspective. 

\section*{Acknowledgements}\noindent
The research conducted by L. Waldorp has been supported by a grant from the
ERC (ERC project 101053880) and by the “New Science of Mental Disorders”, the Dutch
Research Council and the Dutch Ministry of Education, Culture and Science (NWO), grant number 024.004.016. The research by M. Marsman was supported by the Dutch Research Council (VIDI). 

%---------------------------------------------------------------------------
%\bibliography{references}

\begin{thebibliography}{}

\bibitem[Besag, 1974]{Besag:1974}
Besag, J. (1974).
\newblock Spatial interaction and the statistical analysis of lattice systems.
\newblock {\em Journal of the Royal Statistical Society. Series B
  (Methodological)}, 36(2):192--236.

\bibitem[Bork et~al., 2022]{Bork:2022}
Bork, R.~v., Rhemtulla, M., Sijtsma, K., and Borsboom, D. (2022).
\newblock A causal theory of error scores.
\newblock {\em Psychological Methods}.

\bibitem[Castillo et~al., 1998]{Castillo:1998}
Castillo, E., Ferrandiz, J., and Sanmartin, P. (1998).
\newblock Marginalizing in undirected graphs and hypergraphs.
\newblock In {\em Proceedings of the Fourteenth conference on Uncertainty in
  Artificial Intelligence}, pages 69--78. Morgan Kaufmann Publishers.

\bibitem[Cipra, 1987]{Cipra:1987}
Cipra, B. (1987).
\newblock An introduction to the ising model.
\newblock {\em The American Mathematical Monthly}, 94(10):937--959.

\bibitem[Cover and Thomas, 2006]{Cover:2006}
Cover, T. and Thomas, J. (2006).
\newblock {\em Elements of information theory}.
\newblock Wiley and Sons, 2nd edition.

\bibitem[Cowell et~al., 1999]{Cowell:1999}
Cowell, R., Dawid, A., Lauritzen, S., and Spiegelhalter, D. (1999).
\newblock {\em Probabilistic networks and expert systems}.
\newblock Springer.

\bibitem[Cronbach et~al., 1963]{Cronbach:1963}
Cronbach, L.~J., Rajaratnam, N., and Gleser, G.~C. (1963).
\newblock Theory of generalizability: A liberalization of reliability theory.
\newblock {\em British Journal of Statistical Psychology}, 16(2):137--163.

\bibitem[De~Chiusole et~al., 2024]{Chiusole:2024}
De~Chiusole, D., Granziol, U., Spoto, A., and Stefanutti, L. (2024).
\newblock Reliability of a probabilistic knowledge structure.
\newblock {\em Behavior Research Methods}, 56(7):8022--8037.

\bibitem[Ellis and Junker, 1997]{Ellis:1997}
Ellis, J. and Junker, B. (1997).
\newblock Tail-measurablility in monotone latent varible models.
\newblock {\em Psychometrika}, 62(4):495--523.

\bibitem[Garban and Steif, 2014]{Garban:2014}
Garban, C. and Steif, J. (2014).
\newblock {\em Noise Sensitivity of Boolean Functions and Percolation}.
\newblock Cambridge University Press, Cambridge.

\bibitem[Goldberg et~al., 1988]{Goldberg:1988}
Goldberg, D., Bridges, K., Duncan-Jones, P., and Grayson, D. (1988).
\newblock Detecting anxiety and depression in general medical settings.
\newblock {\em British Medical Journal}, 297(6653):897--899.

\bibitem[Haslbeck and Waldorp, 2015]{Haslbeck:2015}
Haslbeck, J. and Waldorp, L.~J. (2015).
\newblock Structure estimation for mixed graphical models in high-dimensional
  data.
\newblock {\em arXiv preprint arXiv:1510.05677}.

\bibitem[Hastie et~al., 2015]{Hastie:2015}
Hastie, T., Tibshirani, R., and Wainwright, M. (2015).
\newblock {\em Statistical learning with sparsity: the lasso and
  generalizations}.
\newblock CRC Press.

\bibitem[Holland and Rosenbaum, 1986]{Holland:1986b}
Holland, P. and Rosenbaum, P. (1986).
\newblock Conditional association and unidimensionality in monotone latent
  variable models.
\newblock {\em The Annals of Statistics}, 14(4):1523--1543.

\bibitem[Hyv{\"a}rinen, 2006]{Hyvarinen:2006}
Hyv{\"a}rinen, A. (2006).
\newblock Consistency of pseudolikelihood estimation of fully visible boltzmann
  machines.
\newblock {\em Neural Computation}, 18(10):2283--2292.

\bibitem[Javanmard and Montanari, 2014]{Javanmard:2014}
Javanmard, A. and Montanari, A. (2014).
\newblock Confidence intervals and hypothesis testing for high-dimensional
  regression.
\newblock Technical report, arXiv:1306.317.

\bibitem[Junker, 1993]{Junker:1993}
Junker, B.~W. (1993).
\newblock Conditional association, essential independence and monotone
  unidimensional item response models.
\newblock {\em The Annals of Statistics}, 21(3):1359--1378.

\bibitem[Kelly, 1988]{Kelly:1988}
Kelly, J.~S. (1988).
\newblock {\em Social choice theory: An introduction}.
\newblock Springer Science \& Business Media.

\bibitem[Lauritzen, 1996]{Lauritzen96}
Lauritzen, S. (1996).
\newblock {\em Graphical Models}.
\newblock Oxford University Press, Oxford.

\bibitem[Loh and Wainwright, 2013]{Loh:2013}
Loh, P.-L. and Wainwright, M.~J. (2013).
\newblock Structure estimation for discrete graphical models: Generalized
  covariance matrices and their inverses.
\newblock {\em The Annals of Statistics}, 41(6):3022--3049.

\bibitem[Lord and Novick, 1968]{Lord:1968}
Lord, F.~M. and Novick, M.~R. (1968).
\newblock {\em Statistical theories of mental test scores}.
\newblock IAP.

\bibitem[Maris and Van~der Maas, 2012]{Maris:2012}
Maris, G. and Van~der Maas, H. (2012).
\newblock Speed-accuracy response models: Scoring rules based on response time
  and accuracy.
\newblock {\em Psychometrika}, 77(4):615--633.

\bibitem[Marsman et~al., 2018]{Marsman:2018}
Marsman, M., Borsboom, D., Kruis, J., Epskamp, S., van Bork, R., Waldorp, L.,
  Maas, H. v.~d., and Maris, G. (2018).
\newblock An introduction to network psychometrics: Relating ising network
  models to item response theory models.
\newblock {\em Multivariate Behavioral Research}, 53(1):15--35.

\bibitem[May, 1952]{May:1952}
May, K.~O. (1952).
\newblock A set of independent necessary and sufficient conditions for simple
  majority decision.
\newblock {\em Econometrica: Journal of the Econometric Society}, pages
  680--684.

\bibitem[Mossel et~al., 2010]{Mossel:2010}
Mossel, E., O'Donnell, R., and Oleszkiewicz, K. (2010).
\newblock Noise stability of functions with low influences: invariance and
  optimality.
\newblock {\em Annals of Mathematics}, 17(1):295--341.

\bibitem[Nguyen, 2017]{Nguyen:2017}
Nguyen, H. (2017).
\newblock Near universal consistency of the maximum pseudolikelihood estimator
  for discrete models.
\newblock {\em Annals of Statistics}, 2(2):22--23.

\bibitem[O'Donnell, 2014]{ODonnell:2014}
O'Donnell, R. (2014).
\newblock {\em Analysis of boolean functions}.
\newblock Cambridge University Press.

\bibitem[Plischke and Bergersen, 2006]{Plischke:1994}
Plischke, M. and Bergersen, B. (2006).
\newblock {\em Equilibrium statistical physics}.
\newblock World Scientific Publishing Company, 3rd edition.

\bibitem[P{\"o}tscher and Schneider, 2009]{Potscher:2009c}
P{\"o}tscher, B.~M. and Schneider, U. (2009).
\newblock On the distribution of the adaptive lasso estimator.
\newblock {\em Journal of Statistical Planning and Inference},
  139(8):2775--2790.

\bibitem[{R Development Core Team}, 2012]{R:2012}
{R Development Core Team} (2012).
\newblock {\em R: A Language and Environment for Statistical Computing}.
\newblock R Foundation for Statistical Computing, Vienna, Austria.
\newblock {ISBN} 3-900051-07-0.

\bibitem[Rasch, 1960]{Rasch:1960}
Rasch, G. (1960).
\newblock Studies in mathematical psychology: I. probabilistic models for some
  intelligence and attainment tests.

\bibitem[Ravikumar et~al., 2010]{Ravikumar:2010}
Ravikumar, P., Wainwright, M., and Lafferty, J. (2010).
\newblock High-dimensional ising model selection using $\ell_1$-regularized
  logistic regression.
\newblock {\em The Annals of Statistics}, 38(3):1287--1319.

\bibitem[Rosenbaum, 1984]{Rosenbaum:1984}
Rosenbaum, P.~R. (1984).
\newblock Testing the conditional independence and monotonicity assumptions of
  item response theory.
\newblock {\em Psychometrika}, 49:425--435.

\bibitem[Schwartzberg, 2008]{Schwartzberg:2008}
Schwartzberg, M. (2008).
\newblock Voting the general will: Rousseau on decision rules.
\newblock {\em Political Theory}, 36(3):403--423.

\bibitem[Sijtsma et~al., 2024]{Sijtsma:2024}
Sijtsma, K., Ellis, J.~L., and Borsboom, D. (2024).
\newblock Recognize the value of the sum score, psychometrics' greatest
  accomplishment.
\newblock {\em Psychometrika}, 89(1):84--117.

\bibitem[Sijtsma and Molenaar, 1987]{Sijtsma:1987}
Sijtsma, K. and Molenaar, I.~W. (1987).
\newblock Reliability of test scores in nonparametric item response theory.
\newblock {\em Psychometrika}, 52(1):79--97.

\bibitem[van Borkulo et~al., 2014]{Borkulo:2014}
van Borkulo, C.~D., Borsboom, D., Epskamp, S., Blanken, T.~F., Boschloo, L.,
  Schoevers, R.~A., and Waldorp, L.~J. (2014).
\newblock A new method for constructing networks from binary data.
\newblock {\em Scientific Reports}, 4:5918.

\bibitem[van~de Geer et~al., 2013]{Geer2013}
van~de Geer, S., B{\"u}hlmann, P., and Ritov, Y. (2013).
\newblock On asymptotically optimal confidence regions and tests for
  high-dimensional models.
\newblock {\em arXiv preprint arXiv:1303.0518}.

\bibitem[van~de Geer et~al., 2014]{Geer:2014}
van~de Geer, S., B{\"u}hlmann, P., Ritov, Y., and Dezeure, R. (2014).
\newblock On asymptotically optimal confidence regions and tests for
  high-dimensional models.
\newblock {\em The Annals of Statistics}, 42(3):1166--1202.

\bibitem[Van~de Geer, 2008]{Geer:2008}
Van~de Geer, S.~A. (2008).
\newblock High-dimensional generalized linear models and the lasso.
\newblock {\em The Annals of Statistics}, pages 614--645.

\bibitem[van~der Linden, 1980]{Linden:1980}
van~der Linden, W.~J. (1980).
\newblock Decision models for use with criterion-referenced tests.
\newblock {\em Applied Psychological Measurement}, 4(4):469--492.

\bibitem[van~der Linden, 1987]{Linden:1987}
van~der Linden, W.~J. (1987).
\newblock The use of test scores for classification decisions with threshold
  utility.
\newblock {\em Journal of educational statistics}, 12(1):62--75.

\bibitem[Van~der Linden, 2018]{Linden:2018}
Van~der Linden, W.~J. (2018).
\newblock {\em Handbook of Item Response Theory: Volume three: Applications},
  volume~3.
\newblock CRC press.

\bibitem[van~der Linden and Hambleton, 2013]{Linden:2013}
van~der Linden, W.~J. and Hambleton, R.~K. (2013).
\newblock {\em Handbook of modern item response theory: Volume one}, volume~1.
\newblock CRC Press.

\bibitem[Wainwright, 2019]{Wainwright:2019}
Wainwright, M.~J. (2019).
\newblock {\em High-dimensional statistics: A non-asymptotic viewpoint},
  volume~48.
\newblock Cambridge University Press.

\bibitem[Wainwright and Jordan, 2008]{Wainwright:2008}
Wainwright, M.~J. and Jordan, M.~I. (2008).
\newblock Graphical models, exponential families, and variational inference.
\newblock {\em Foundations and Trends in Machine Learning}, 1(1-2):1--305.

\bibitem[Waldorp et~al., 2019]{Waldorp:2019}
Waldorp, L., Marsman, M., and Maris, G. (2019).
\newblock Logistic regression and ising networks: prediction and estimation
  when violating lasso assumptions.
\newblock {\em Behaviormetrika}, 46(1):49--72.

\bibitem[Yang et~al., 2012]{Yang:2012}
Yang, E., Allen, G., Liu, Z., and Ravikumar, P.~K. (2012).
\newblock Graphical models via generalized linear models.
\newblock In {\em Advances in Neural Information Processing Systems}, pages
  1358--1366.

\bibitem[Yang et~al., 2013]{Yang:2013}
Yang, E., Ravikumar, P., Allen, G.~I., and Liu, Z. (2013).
\newblock On graphical models via univariate exponential family distributions.
\newblock {\em arXiv preprint arXiv:1301.4183}.

\end{thebibliography}

%---------------------------------------------------------------------------
%---------------------------------------------------------------------------
\appendix 
\vspace{2em}\noindent
\textbf{Appendix}

\section{Proofs of main theorems and propositions}\label{sec:proofs}\noindent

%---------------------------------------------------------------------------
\subsection{Kullback-Leibler divergence between $\mathbb{P}_\pi$ and $\mathbb{P}$}\label{proof-sec:kl-divergence}
\noindent
{\bf Lemma \ref{lem:kl-divergence-probability}}
Let $\Prob$ be the joint Ising probability (\ref{eq:ising-joint-probability}) and $\Prob_{\pi}$ be the product of conditionals (\ref{eq:joint-measure}). Then the Kullback-Leibler divergence is
\begin{align*}
D_{\rm KL}(\Prob \mid\mid \Prob_{\pi})=\log \frac{Z_{V}}{\prod_{i\in V}Z_{i}}
\end{align*}
%

%----------------------------------------------------------------
\begin{proof}
The Kullback-Leibler divergence is 
\begin{align*}
D_{\rm KL}(\Prob\mid\mid \Prob_{\pi}) = %\sum_{x\in \{-1,1\}^{n}}\Prob(x)\log \frac{\Prob(x)}{\Prob_{\pi}(x)}= 
 	\sum_{x\in \{-1,1\}^{n}}\Prob(x)\log \Prob(x) - \sum_{x\in \{-1,1\}^{n}}\sum_{i\in V}\Prob(x)\log \Prob_{i}(x)
\end{align*}
By Lemma \ref{lem:conditional-product-measure}, where we use the rescaled univariate conditionals with interaction parameters $\theta_{ij}/2$, we have 
\begin{align*}
\Prob(x) = \frac{\prod_{i\in V}Z_{i}(x_{\partial i})}{Z_{V}} \prod_{i\in V}\frac{1}{Z_{i}(x_{\partial i})}\exp\left( \xi_{i}x_{i} + \frac{1}{2}\sum_{j\in \partial i}\theta_{ij}x_{i}x_{j}\right)
\end{align*}
And so we see that we get from the rescaled product of conditionals to the joint distribution of the Ising probability. Plugging in the KL divergence for each $i$ the univariate conditional
\begin{align*}
\Prob_{i}(x) = \frac{1}{Z_{i}(x_{\partial i})}\exp\left( \xi_{i}x_{i} + \frac{1}{2}\sum_{j\in \partial i}\theta_{ij}x_{i}x_{j}  \right)
\end{align*}
gives the result. 
\end{proof}

\subsection{Orthonormal base for $p$-biased and dependent variables}\label{proof-sec:p-biased-basis}
\noindent
{\bf Proposition \ref{lem:p-biased-basis}} 
Let $\phi_S(X)$ be the transformation in (\ref{eq:z-transform-phi})  for $X_i$ $p$-biased for $i$ and dependent. Using the pseudo-likelihood approach (\ref{eq:joint-measure}), we obtain for subset $S$ and $T$ of $V$
\begin{align*}
\E_{\pi}(\phi_{S}(X)\phi_{T}(X)) 
=
\begin{cases}
1	&\text{ if } S=T\\
0	&\text{ if } S\ne T\\ 
\end{cases}
\end{align*}
\begin{proof}
Using the definition of $\phi_S$
\begin{align*}
\E_{\pi}(\phi_{S}(X)\phi_{T}(X)) = \E_{\pi}\left(\prod_{i\in S}\phi(X_{i})\prod_{i\in T}\phi(X_{i})\right)
\end{align*}
We next apply the product of expectations in (\ref{eq:expectation-product}) to get that for any $i\in S$ and $j\in T$ such that $i\ne j$,  $\mathbb{E}_\pi(\phi(X_{i})\phi(X_{j}))=\mathbb{E}_{p_i}(\phi(X_{i}))\mathbb{E}_{p_j}(\phi(X_{j}))$. 

If $S=T$ then all terms are $\E_{\pi}\phi(X_{i})^{2}$, and this equals the variance, and $\text{var}_{p_{i}}(\phi(X_{i}))=1$ for all $i\in S$. 

If $S\ne T$ then there is at least one $i\notin S\cap T$ and, if we suppose $i\notin T$ but $i\in S$, then
\begin{align*}
\E_{p_{i}}\phi(X_{i})\E_{\pi\backslash p_{i}}\left(\prod_{j\in S\backslash \{i\}}\phi(X_{j})\prod_{j\in T}\phi(X_{j})\right)
\end{align*}
where  $\E_{p_{i}}\phi(X_{i})$ equals 0. Hence, the only terms in the Fourier expansion that remain are when $S=T$. And so, we obtain (\ref{eq:basis-function-orthogonal-p-biased}).
\end{proof}
%

%---------------------------------------------------------------------------
\subsection{Mean, variance and covariance of $p$-biased and dependent $X_i$}\label{p-biased moments}
\noindent
{\bf Proposition \ref{lem:p-biased-mean-variance-covariance}} 
Let $f$ be a Boolean function and let the $X_{i}$ be biased and dependent in that $\E_{p_{i}}(X_{i})=2p_{i}-1$ for each $i$, based on the conditional distributions obtained with the Ising model. Then the mean, variance and covariance are, respectively,
\begin{align*}
\E_{\pi} (f(X)) &= \E_{\pi}\left( \sum_{S\subseteq V}\hat{f}^{\pi}(S)\phi_{S}(X)\right) =  \hat{f}^{\pi}(\varnothing)
\end{align*}
and
\begin{align*}
\text{var}_{\pi}(f) = \E_{\pi} f(X)^{2} -(\E_{\pi} f(X))^{2}=  \sum_{S\neq\varnothing}\hat{f}^{\pi}(S)^{2}
\end{align*}
and
\begin{align*}
\text{cov}_{\pi}(f) = \E_{\pi} f(X)g(X) -\E_{\pi} f(X)\E_{\pi} g(X)=  \sum_{S\neq\varnothing}\hat{f}^{\pi}(S)\hat{g}^{\pi}(S)
\end{align*}
\begin{proof}
The proof is similar for the variance and the covariance, since for the covariance $\text{cov}_{\pi}$ both $f$ and $g$ are decomposed in terms of the functions $\phi_{S}$; so we only prove the variance here. 

The mean is obvious since the only term of $\E_{\pi}(\phi_{S}(X))$ that is non-zero is when $S=\varnothing$. For the variance, we have
\begin{align*}
 \E_{\pi} f(X)^{2} 
 %= \E_{\pi}\left( \sum_{S\subseteq V}\hat{f}^{\pi}(S)\phi_{S}(x)\sum_{T\subseteq V}\hat{f}^{\pi}(T)\phi_{T}(x) \right)
 = \sum_{S\subseteq V}\sum_{T\subseteq V}\hat{f}^{\pi}(S)\hat{f}^{\pi}(T)\E_{\pi}(\phi_{S}(X)\phi_{T}(X))
\end{align*}
By Lemma \ref{lem:p-biased-basis} the only terms in the Fourier expansion that remain are when $S=T$, and so we obtain the variance.
\end{proof}
%

%
%%---------------------------------------------------------------------------
%\begin{lemma}\label{lem:influence-upper-bound}
%Let $f:\{-1,1\}^{n}\to \{-1,1\}$ be monotone and transitive-symmetric (properties (a) and (e) of Section \ref{sec:boolean-functions}). Then 
%%
%\begin{align*}
%\mathbb{I}_{i}(f) \le \frac{1}{\sqrt{n}}
%\end{align*}
%%
%\end{lemma}
%%
%\begin{proof}
%By the property of transitive-symmetry we have that $\hat{f}(i)=\hat{f}(i')=\hat{f}(1)$, which holds for any $i\in V$. Thus by monotonicity from Lemma \ref{lem:influence-fourier-s1} $\mathbb{I}_{i}(f)=\hat{f}(i)$ for all $i\in V$. By Parseval 
%%
%\begin{align*}
%\sum_{S\subseteq V}\hat{f}(S)^{2}=1\le \sum_{i=1}^{n}\hat{f}(i)^{2}=n\hat{f}(i)^{2}
%\end{align*}
%%
%which gives the result. 
%\end{proof}
%%

\subsection{Bound on influence}\label{proof-sec:bound-influence}
\noindent
{\bf Proposition \ref{prop:bound-influence}}
Let $f$ be a Boolean monotone function and assume transitive symmetry (e) from Section \ref{sec:boolean-functions} for $p$-biased and dependent $X_i$ given the measure $\mathbb{P}_\pi$ in (\ref{eq:joint-measure}). Then 
\begin{align}
\mathbb{I}_{i}^{\pi}(f)\le \frac{\text{sd}_{\pi}(f)}{\sigma_{i}\sqrt{n}}
\end{align}
\begin{proof}
We have from Lemma \ref{lem:p-biased-mean-variance-covariance} that the variance of a Boolean function can be represented as $\text{var}_{\pi}(f)=\sum_{S\ne \varnothing} \hat{f}^{\pi}(S)^{2}$. Clearly this is larger than taking only the subsets $S=\{i\}$, and so, $\text{var}_{\pi}(f)\ge \sum_{i\in V} \hat{f}^{\pi}(i)^{2}$. Assuming transitive-symmetry (e), we know that nodes are exchangeable, and so $\hat{f}^{\pi}(i)=\hat{f}^{\pi}(j)$ for any nodes $i$ and $j$. And so, if we assume that $f$ is monotone, we have that 
\begin{align*}
\text{var}_{\pi}(f)\ge \sum_{i\in V}\hat{f}^{\pi}(i)^{2}=n\hat{f}^{\pi}(i)^{2}
\end{align*}
Hence, we obtain the upper bound for the influence of item $i$.
\end{proof}

%---------------------------------------------------------------------------
\subsection{Equal degree and coefficients Ising graph implies equal influence}\label{proof-sec:equal-degree-influence}
\noindent
{\bf Proposition \ref{prop:influence-same-degree-same}} 
Let the graph $G$ be induced by a regular (equal degree nodes) Ising model (\ref{eq:ising-joint-probability}) with equal interaction and threshold parameters. Then the decision function $f$ will have equal influences for all items. \\

\begin{proof}
Suppose that the probabilities for each node are the same, $\Prob (x_{i}\mid x_{\partial i}) = p$ for all $i\in V$. We need to show that (a) we have the same conditional probabilities for any node, and (b) that this leads to the same influence for each node. 

We start with (b). Assuming the same probabilities for each node, we have that $\mu_{i}=\mu$ and $\sigma_{i}=\sigma$ for all $i\in V$. And so the functions $\phi(x_{i})$ are all the same, given a value $x_{i}$. Then we find for the influence 
\begin{align*}
\E_{\pi}(f(X)\phi(X_{i})) = \E_{p^{n-1}} \left(f(X)\phi(1)p - f(X^i)\phi(-1)(1-p)\right)
\end{align*}
where $p^{n-1}$ refers to the sequence of $n-1$ equal probabilities $(p,p,\ldots,p)$.
Recall that $\phi(1)=\sqrt{(1-p)/p}$ and $\phi(-1)=\sqrt{p/(1-p)}$, so that $p\phi(1)=(1-p)\phi(-1)=\sqrt{p(1-p)}$ and $\sigma=2\sqrt{p(1-p)}$. Then we obtain 
\begin{align*}
\E_{\pi}(f(X)\phi(X_{i}))= \E_{p^{n-1}}\left( f(X) - f(X^i)\right) \frac{1}{2}\sigma
\end{align*}
Recall that the part of the right hand side in the expectation operator equals the discrete differential operator $D_{i}f$, and its expectation $\E_{\pi}(D_{i}f)$ is by definition the influence. Because all probabilities $\Prob (x_{i}\mid x_{\partial i}) = p$ are equal, we obtain each time the same expectation. 

We next consider (a), the equality of the probabilities $p_{i}$ for $i\in V$. 
Because we assumed that the graph is regular, we see that the size of the boundary sets $|\partial i|=r$ is equal for all $i\in V$. By assuming additionally that the threshold and interaction parameters are all the same, i.e., $\xi=\xi_{i}$ for all nodes and $\beta=\beta_{ij}$ for all edges, we obtain that the probabilities, defined by the function $\xi+\beta\sum_{k\in \partial i} X_{ik}$, with the same size sets $\partial i$ for all $i$, are equal up to differences in $S_{r}=\sum_{k\in \partial i} X_{ik}$. However, $S_{r}$ contains for each node the same number of $X_{ik}$ that have the same conditional distribution. Hence, we have a spatially stationary (shift invariant) process in which the probabilities are the same across the graph. 
%By consequence the The sum $s_{r}$ of $r$ variables is determined by the probabilities of each $X_{ik}$. Each $X_{ik}$ has expectation $\mu_{i}=\Prob_{p_{i}}(X_{i}=1\mid x_{\partial i})-\Prob_{p_{i}}(X_{i}=-1\mid x_{\partial i})=2p_{i}-1$ and variance $\sigma^{2}_{i}=1-\mu_{i}^{2}$. 
%Using Hoeffding's inequality  \citep[e.g.,][Theorem 2.2.6]{Vershynin:2018}, we obtain that with high probability that we can replace $s_{r}$ by $\sum_{j\in \partial i}\mu_{j}$. To see this, let $\delta = \exp(-t^{2}/(2r))$ so that $t=r\sqrt{2\log(2/\delta)}$. Then Hoeffding's inequality tells us that 
%%
%\begin{align*}
%\Prob_{\pi}\left( \left| \sum_{j\in \partial i} (X_{j}-\mu_{j})\right|  \le r\sqrt{2\log(\tfrac{2}{\delta})} \right) \ge 1-\delta
%\end{align*}
%%
%Hence, we see that we can replace $s_{r}$ by $\sum_{j\in \partial i}\mu_{j}$  with probability $1-\delta$. 
%
%We assume that $p_{i}\in (p-\tfrac{1}{2}\varepsilon,p+\tfrac{1}{2}\varepsilon)$ for all $i\in V$. Then $\mu_{i}$ is wrong by at most $\varepsilon$, because $\mu_{i}\in (\mu-\varepsilon,\mu+\varepsilon)$ and $\sigma_{i}$ is in $(\sigma-\varepsilon,\sigma+\varepsilon)$. Consequently, we obtain that the expectation is within $\varepsilon^{n}$ of the true expectation.  If $\varepsilon$ is small, then the probabilities will be almost identical. 
\end{proof}

\subsection{Probabilistic measurement error aligns with classical test theory}\label{proof-sec:alignment-ctt}
\noindent
{\bf Lemma \ref{lem:assumptions-classic-tt}} 
The assumptions of classical test theory, as stated in \citet[][Theorem 2.7.1]{Lord:1968}, are satisfied by random variables $X_i$ with independent contaminated $Y_i$ and $Z_i$, using the set-up with measurement error defined by (\ref{eq:measurement-error}), i.e.,  
\begin{itemize}
\item[(i)] the expectation of the error is 0: $\E_{p_i,\rho}(Y_{i}-\mu_{\rho}(X_{i}))=0$,
\item[(ii)] the correlation between the error and the true score is 0: $\text{\rm cor}_{p_{i},\rho}(Y_{i}-\mu_{\rho}(x_{i}),\rho X_{i})=0$
\item[(iii)] the correlation between the error and the true score from another measurement is 0: $\text{\rm cor}_{p_{i},\rho}(Y_{i}-\mu_{\rho}(x_{i}),\rho X_{i})=0$ 
\item[(iv)] the correlation between distinct error measurements is 0: 
\end{itemize}

\begin{proof}
(i) Fix the uncentered covarince between $X_{i}$ and $Y_{i}$, for any $i\in V$, to $\rho$. Let the error be $Y_{i}-\mu_{\rho}(x_{i})=Y_i-\mathbb{E}(Y_i\mid X_i)=Y_i-\rho X_i$. Then 
\begin{align}
\E_{\rho}(Y_{i}-\mu_{\rho}(X_{i}))=X_{i}\frac{1}{2}(1+\rho) - X_{i}\frac{1}{2}(1-\rho) - \rho X_{i} =0
\end{align}

(ii) Using (i) we find that the correlation between the error and the true score is 0, because
\begin{align}
\text{cov}_{p_{i},\rho}(Y_{i}-\mu_{\rho}(X_{i}),\rho X_{i}) = \rho\E_{p_{i},\rho}(Y_{i}X_{i})-\rho^{2}\E_{p_{i}}(X_{i}^{2})=0
\end{align}
since $\E_{p_{i},\rho}(Y_{i}X_{i})=\rho\E_{p_{i}}(X_{i}^{2})$. 

(iii) Let $Z_i$ be the contaminated score obtained from $X_i$. Then its true score is $\rho X_i$. Hence, by (ii) we obtain that the correlation between the error $Y_i-\mu_\rho(X_i)$ and the true score $\rho X_i$ of $Z_i$, is 0.

(iv) To show that the errors of two different, independent measurements are zero-correlated, we use $Y$ and $Z$ obtained from the same original $X$ with probability $\tfrac{1}{2}(1+\rho)$ that the value $Y_i$ and $Z_i$ are equal to $X_i$, each with the same parameter $\rho$. And so the true score of both $Y_i$ and $Z_i$ is $\mu_\rho(X_i)=\rho X_i$. Then we obtain for the covariance between $Y_i-\rho X_i$ and $Z_i-\rho X_i$
\begin{align*}
&\text{cov}_{p_i,\rho}(Y_i-\rho X_i,Z_i-\rho X_i) 
 \\&\quad = \E_{p_i,\rho}(Y_iZ_i)-\rho\E_{p_i,\rho}(Y_iX_i)-\rho\E_{p_i,\rho}(Z_iX_i)+\rho^2\E_{p_i,\rho}(X_i^2)
\\&\quad  =2\rho^2\E_{p_i,\rho}(X_i^2)- 2\rho^2\E_{p_i,\rho}(X_i^2)=0
\end{align*}
because of independence between $Y_i$ and $Z_i$, $\mathbb{E}(Y_iZ_i)=\mathbb{E}(Y_i)\mathbb{E}(Z_i)=\rho^2X_i^2$. This was to be proved. 
\end{proof}

%---------------------------------------------------------------------------
\subsection{Reliability of $Y$ and $X$}\label{proof-sec:reliability}
\noindent
{\bf Theorem \ref{thm:reliability}} 
Given $Y_i$ subjected to measurement error (\ref{eq:measurement-error}) with probability $\tfrac{1}{2}(1-\rho)$ of obtaining $-X_i$, the correlation between $Y_i$ and $X_i$ is 
\begin{align}
%\text{cor}_{p_{i},\rho}(X_{i},Y_{i}) =\E_{p_{i},\rho}\left(\frac{X_{i}-\mu_{i}}{\sigma_{i}}\right)\left(\frac{Y_{i}-\rho\mu_{i}}{\sigma_{i,\rho}}\right)
\text{cor}_{p_{i},\rho}(X_{i},Y_{i}) = \E_{p_{i},\rho}(\phi(X_{i})\phi_{\rho}(Y_{i}))=\rho\frac{\sigma_{i}}{\sigma_{i,\rho}}
\end{align}
And the reliability is
\begin{align}
\text{cor}_{p_i,\rho}^2(X_i,Y_i)=\rho^2\frac{\sigma_{i}^2}{\sigma_{i,\rho}^2}
\end{align}
\begin{proof}
We first note that $\E_{p_{i},\rho}(Y_{i})=\rho \mu_{i}$ and $\text{var}_{p_{i},\rho}(Y_{i}^{2})=1-\rho^{2}\mu_{i}^{2}$. Then 
\begin{align*}
\phi^{\rho}(Y) = \frac{Y_{i}-\rho\mu_{i}}{\sqrt{1-\rho^{2}\mu_{i}^{2}}}
\end{align*}
so that $\E_{p_{i},\rho}(\phi^{\rho}(Y_{i}))=0$ and $\E_{p_{i},\rho}(\phi^{\rho}(Y_{i})^{2})=1$.
We have that the expectation of $\phi(X_{i})\phi^{\rho}(Y_{i})$, with respect to the measure $\Prob_{\rho}$ and then with respect to $\Prob_{p_{i}}$, is
\begin{align*}
\E_{p_{i},\rho}(\phi(X_{i})\phi^{\rho}(Y_{i})) = \frac{1}{2}(1+\rho)\E_{p_{i}}(\phi(X_{i})\phi^{\rho}(X_{i}))+ \frac{1}{2}(1-\rho)\E_{p_{i}}(\phi(X_{i})\phi^{\rho}(-X_{i}))
\end{align*}
The expectations on the right hand side are, using that $\mathbb{E}_{p_i}(\phi(X_i)^2)=\sqrt{1-\mu_i^2}$ 
\begin{align*}
\E_{p_{i}}(\phi(X_{i})\phi^{\rho}(X_{i}))=\frac{\sqrt{1-\mu_{i}^{2}}}{\sqrt{1-\rho^{2}\mu_{i}^{2}}}
	\quad\text{and}\quad
\E_{p_{i}}(\phi(X_{i})\phi^{\rho}(-X_{i})) = -\frac{\sqrt{1-\mu_{i}^{2}}}{\sqrt{1-\rho^{2}\mu_{i}^{2}}} 
\end{align*}
which is the ratio of standard deviations, $\sigma_{i}$ for $X_i$ in the nominator and $\sigma_{i,\rho}$ for $Y_i$ in the denominator. Putting this together with the expectation using $\mathbb{P}_\rho$ above, we obtain the correlation, and the square for the reliability. 
\end{proof}
%

%------------------------
\subsection{Fully connected graph, conditional association, and vanishing conditional dependence }\label{proof-sec:vanishing-conidtional-dependence}
\noindent
{\bf Proposition \ref{cor:vanishing-conidtional-dependence}} 
Let $X$ be the random variables induced by the Ising probability in (\ref{eq:ising-joint-probability}) with respect to $G=(V,E)$ with $|V|\to \infty$ and positive connectivity parameters. Then $X$ is conditionally associated and has the property of vanishing conditional dependence if and only if the graph $G$ is fully connected. 
\\

\begin{proof}
Assume that graph $G$ is fully connected and has positive connectivity parameters. Then any correlation between two nodes in G is positive \citep{Plischke:1994}. And by Theorem 1 in \citet[][]{Castillo:1998} there exists a single (continuous) variable $\theta$ outside $G$ such that (a) all nodes are connected to $\theta$, (b) the marginal is a fully connected graph, and (c) conditioning on $\theta$ renders the graph $G$ empty. An example of such a variable $\theta$ such that the marginal on $G$ over $\theta$ is the Ising model, is given in \citet[][]{Marsman:2018}. By Theorem 5 in \citet{Ellis:1997} it then follows that the latent variable representation is equivalent to $X$ being conditionally associated and has vanishing conditional dependence. 

Now assume that $X$ is conditionally associated and has vanishing conditional dependence. Then by the equivalence in Theorem 5 in \citet{Ellis:1997}, it follows that there is a unidimensional latent variable representation, that is also monotone such that conditioning on this variable renders $G$ empty. By Theorem 1 in \citet{Castillo:1998}, the marginal is fully connected with positive connectivity parameters. 
\end{proof}

%------------------------
\subsection{Conditional association in a connected graph}\label{proof-sec:conidtional-association}
\noindent
{\bf Theorem \ref{thm:conditional-association}}
Let $G$ be a graph with nodes $X$ that follows the Ising probability distribution (\ref{eq:ising-joint-probability}) with equal external field $\xi$ and positive connectivity parameters $\theta_{ij}$ for all $i,j\in V$. Furthermore, let $h:\{-1,1\}^{k}\to D$, with $D\subset \mathbb{Z}$ be some discrete outcome function Then: 
\begin{itemize}
\item[(i)] If $G$ is disconnected (has two or more components), then $X$ defined on $G$ is not conditionally associative. 
\item[(ii)] If $G$ is connected, and the Boolean functions $f$ and $g$ are LTFs with coefficients $(a_i,i\in V)$ and $(b_i,i\in V)$, respectively, such that $\sum_{i\in K} a_i b_i\sigma_i^2\ge 0$, then $X$ is conditionally associative. 
%\item[(3)] If $G$ is fully connected with positive interaction parameter, then $X$ defined on $G$ is conditionally associative. 
\end{itemize}
\begin{proof}
We show here that assuming a graph $G$ induced by the Ising model implies conditional associativity only when it is positively connected (i.e., has positive interaction parameters and has one component). We fix $h(x_L)=c$.

Because $f$ and $g$ are monotone, we can use the Fourier expansion $f(x)=\sum_{i\in V}\hat{f}^{\pi,\rho}(i)\phi_\rho(x_{i})$ to obtain
\begin{align*}
\text{cov}_{\pi,\rho}(f(X_{K}),g(X_{K})\mid h(x_{L})) = \sum_{i,j\in K}\hat{f}^{\pi,\rho}(i)\hat{g}^{\pi,\rho}(j)\E_{\pi,\rho}(\phi_\rho(X_{i})\phi_\rho(X_{j})\mid h(x_{L}))
\end{align*}
We start with (i) and assume that there is no edge between $K$ and $L$. If $i=j$, then
\begin{align*}
\E_{\pi,\rho}(\phi_\rho(X_{i})^{2}\mid h(x_{L}))= \frac{1}{\Prob_{\pi,\rho}(h(X_{L})=c)}\int\phi_\rho(x_{i})^{2}\mathbbm{1}\{h(x_{L})=c\}d\Prob_{\pi,\rho}(x)
\end{align*}
If $K$ and $L$ are independent, i.e., there is no connection between the partitions $K$ and $L$, then we obtain 
\begin{align*}
\frac{1}{\Prob_{\pi}(h(X_{L})=c)}\int \mathbbm{1}\{h(x_{L})=c\}d\Prob_{\pi}(x_{L})\int\phi_\rho(x_{i})^{2}d\Prob_{\pi}(x_{i})=1
\end{align*}
We clearly have that $\Prob_{\pi,\rho}(h(x_{L})=c)=\int \mathbbm{1}\{h(x_{L})=c\}d\Prob_{\pi,\rho}(x_{L})$, and so the first part equals 1. Furthermore, since $\int\phi_\rho(x_{i})^{2}d\Prob_{\pi,\rho}(x_{i})=\E_{p_{i},\rho}(\phi_\rho(X_{i})^{2})$, is the variance, it $=1$, we obtain that $\E_{\pi,\rho}(\phi_\rho(X_{S})^{2}\mid h(x_{L}))=1$, for any two-way partition $K$ and $L$ of $V$. Furthermore, if $i\ne j$ then orthogonality of the Fourier basis functions (\ref{eq:basis-function-orthogonal-p-biased}) implies that this equals 0. Hence, because there is no effect of node $j$ on node $i$, there is no effect of conditioning on $\{h(X_{L})=c\}$. So, the covariance is independent of the condition $h(X_{L})$ and a disconnected graph implies that $X$ is not conditionally associated.  

For (ii), suppose $K$ and $L$ are connected, i.e., if there is a single edge $(i,j)$ such that $i\in K$ and $j\in L$, then $X_{K}$ and $X_{L}$ are not conditionally independent. Then all $\theta_{k\ell}=0$ whenever $k\in K$ and $\ell\in L$ except for $\theta_{ij}$. And so, for the conditional distribution, we obtain 
\begin{align*}
\Prob(X_{i}=1\mid h(x_{L})=c) = \frac{\exp\left(\theta_{ij}x_{j}\right)\exp\left(\xi + \sum_{t\in K\cap \partial i}\theta_{it}x_{t}\right)}{Z^{p_{i}}(x_{\partial i})}
\end{align*}
Then, if $X_{j}=1$ it will increase the probability that $X_{i}=1$, whenever $\theta_{ij}>0$. Hence, whenever there is a positive connection between the partitions $K$ and $L$,  and $h:\{-1,1\}^{k}\to D\subset \mathbb{Z}$, then 
\begin{align*}
\Prob(X_{i}=1\mid h(x_{L})=c) \ge \Prob(X_{i}=1)
\end{align*}
By consequence, we have for the expectation $\E_{p_{i},\rho}(\phi_\rho(X_{i})^{2}\mid h(x_{L})) $
\begin{align*}
\phi_\rho(1)^2\Prob_{p_i,\rho}(X_i=1\mid h(x_{L})=c) +\phi_\rho(-1)^2 \Prob_{p_i,\rho}(X_i=-1\mid h(x_{L})=c)
\end{align*}
Note that $\phi_\rho(1)^2=(1-\rho\mu_i)/(1+\rho\mu_i)$ and $\phi_\rho(-1)^2=(1+\rho\mu)/(1-\rho\mu_i)$, and that
$\Prob_{p_i,\rho}(X_i=1\mid h(x_{L})=c) = \tfrac{1}{2}(1+\rho)\Prob(X_i=1\mid h(x_{L})=c)$ and $\Prob_{p_i,\rho}(X_i=-1\mid h(x_{L})=c) = \tfrac{1}{2}(1-\rho)\Prob(X_i=-1\mid h(x_{L})=c)$. And we showed above that when $L$ and $K$ are connected
$\Prob(X_{i}=1\mid h(x_{L})=c) \ge \Prob(X_{i}=1)$, so that
\begin{align*}
\E_{p_{i},\rho}(\phi_\rho(X_{i})^{2}\mid h(x_{L}))\ge \E_{p_{i}\rho}(\phi_\rho(X_{i})^{2}) = 1
\end{align*}
So, the conditional covariance increases as a function of $h$ over $X_{L}$ when $K$ and $L$ are connected and all connectivity parameters are positive. 

Next, we need to show that the product of the Fourier coefficients $\hat{f}^{\pi,\rho}(i)\hat{g}^{\pi,\rho}(j)$ is positive. By definition of the Fourier coefficient and assuming an LTF, we find
\begin{align*}
\hat{f}^{\pi,\rho}(i)=\E_{\pi}((a_{0}+a_{1}X_{1}+\cdots +a_{n}X_{n})\phi(X_{i})\mid h(x_L)=c)
\end{align*}
By orthogonality, as before, we only need to consider the case $i=j$. And then 
\begin{align*}
\hat{f}^{\pi,\rho}(i)=\mathbb{E}_{p_i,\rho}(f(X)\phi_\rho(X_i)\mid h(x_L)=c)
\ge 
\hat{f}^{\pi,\rho}(i)=\mathbb{E}_{p_i,\rho}(f(X)\phi_\rho(X_i))
\end{align*}
giving
\begin{align*}
\hat{f}^{\pi,\rho}(i)
%&\ge a_i\mathbb{E}_{p_i,\rho}(X_i\phi_\rho(X_i))
%= a_i \mathbb{E}_{p_i,\rho}(X_i^2/\sigma_i-X_i\mu_i/\sigma_i)\\
\ge a_i[\mathbb{E}_{p_i,\rho}(X_i^2/\sigma_i) - \mu_i^2/\sigma_i]
= a_i\sigma_i
\end{align*}
And so, for the first order coefficients we obtain
\begin{align*}
\text{cov}_{\pi,\rho}(f(X_{K}),g(X_{K})\mid h(x_{L})) = \sum_{i\in K}\hat{f}^{\pi}(i)\hat{g}^{\pi}(i) \ge \sum_{i\in K}a_{i}b_{i}\sigma_{i}^{2}
\end{align*}
which is what was required. Now, assuming that $\sum_{i\in K}a_{i}b_{i}\sigma_{i}^{2}\ge 0$, we obtain conditional association.
\end{proof}
%

%---------------------------------------------------------------------------
\subsection{Covariance of Boolean functions $f(Y)$ and $f(X)$}\label{proof-sec:var-cov-f}
\noindent
{\bf Theorem \ref{lem:variance-covariance-f(x)-f(y)}} 
Let $f:\{-1,1\}^{n}\to \{-1,1\}$ be a Boolean function and for all $i$, $Y_{i}$ are obtained from the experiment where with probability $\tfrac{1}{2}(1+\rho)$, $Y_{i}=X_{i}$ and with probability $\tfrac{1}{2}(1-\rho)$, $Y_{i} =-X_{i}$ for each $i$ independently. Then the variance of $f(Y)$ is 
\begin{align*}
\text{var}_{\pi,\rho}(f(Y)) = \sum_{S\ne \varnothing} \hat{f}^{\pi,\rho}(S)^{2}
\end{align*}
and the covariance between $f(X)$ and $f(Y)$ is 
\begin{align*}
\text{cov}_{\pi,\rho}(f(X),f(Y)) = \sum_{S\ne \varnothing} \omega(S)\rho^{|S|}\hat{f}^{\pi}(S)\hat{f}^{\pi,\rho}(S)
\end{align*}
where $\omega(S)=\prod_{i\in S}\tfrac{\sigma_{i}}{\sigma_{i,\rho}}$.

\begin{proof}
We require the covariance between $f(X)$ and $f(Y)$ and their variances. We start with the variance and then determine the covariance. 
The variance of $f(X)$ is determined in Lemma \ref{lem:p-biased-mean-variance-covariance}. To determine the variance of $f(Y)$ we use the Fourier expansion $f(y)=\sum_{S\subseteq V}\hat{f}^{\pi}(S)\phi_{S}(y)$. We first note that $\E_{p_{i},\rho}(Y_{i})=\rho \mu_{i}$ and $\E_{p_{i},\rho}(Y_{i}^{2})=1-\rho^{2}\mu_{i}^{2}$. Then 
\begin{align*}
\phi^{\rho}(Y) = \frac{Y_{i}-\rho\mu_{i}}{\sqrt{1-\rho^{2}\mu_{i}^{2}}}
\end{align*}
so that $\E_{p_{i},\rho}(\phi^{\rho}(Y_{i}))=0$ and $\E_{p_{i},\rho}(\phi^{\rho}(Y_{i})^{2})=1$.
 We obtain the Fourier coefficients
\begin{align*}
\hat{f}^{\pi,\rho}(S) = \E_{\pi,\rho}(f(Y)\phi^{\rho}_{S}(Y))
\end{align*}
and we have again orthonormality for the Fourier expansion as before (\ref{eq:basis-function-orthogonal-p-biased}). Then, for the variance, we have
\begin{align*}
 \E_{\pi,\rho} f(Y)^{2} 
 %= \E_{\pi}\left( \sum_{S\subseteq V}\hat{f}^{\pi}(S)\phi_{S}(x)\sum_{T\subseteq V}\hat{f}^{\pi}(T)\phi_{T}(x) \right)
 = \sum_{S\subseteq V}\sum_{T\subseteq V}\hat{f}^{\pi}(S)\hat{f}^{\pi}(T)\E_{\pi,\rho}(\phi_{S}^{\rho}(Y)\phi_{T}^{\rho}(Y))
\end{align*}
which is nonzero only if $S=T$. Now we can apply Lemma \ref{lem:p-biased-mean-variance-covariance} for the variance and obtain the result. 

For the covariance we have 
\begin{align*}
\text{cov}^{\pi,\rho}(f(X),f(Y))=\sum_{S,T\subseteq V}\hat{f}^{\pi}(S)\hat{f}^{\pi,\rho}(T)\E_{\pi,\rho}(\phi_{S}(X)\phi^{\rho}_{T}(Y))
\end{align*}
and by orthogonality
\begin{align*}
\E_{\pi,\rho}(\phi_{S}(X)\phi^{\rho}_{S}(Y))=\prod_{i\in S}\E_{\pi,\rho}(\phi(X_{i})\phi^{\rho}(Y_{i}))
\end{align*}
We have that the expectation of $\phi(X_{i})\phi^{\rho}(Y_{i})$, with respect to the measure $\Prob_{\rho}$ and then with respect to $\Prob_{p_{i}}$, is
\begin{align*}
\E_{p_{i},\rho}(\phi(X_{i})\phi^{\rho}(Y_{i})) = \frac{1}{2}(1+\rho)\E_{p_{i}}(\phi(X_{i})\phi^{\rho}(X_{i}))+ \frac{1}{2}(1-\rho)\E_{p_{i}}(\phi(X_{i})\phi^{\rho}(-X_{i}))
\end{align*}
These expectations are
\begin{align*}
\E_{p_{i},\rho}(\phi(X_{i})\phi^{\rho}(Y_{i}))=\frac{\sqrt{1-\mu_{i}^{2}}}{\sqrt{1-\rho^{2}\mu_{i}^{2}}}
	\quad\text{and}\quad
\E_{p_{i},\rho}(\phi(X_{i})\phi^{\rho}(-X_{i})) = -\frac{\sqrt{1-\mu_{i}^{2}}}{\sqrt{1-\rho^{2}\mu_{i}^{2}}} 
\end{align*}
which is the ratio of standard deviations, $\sigma_{i}$ in the nominator and $\sigma_{i}^{\rho}$ in the denominator.
And so, considering the function $\prod_{i\in S} \E_{\pi,\rho}(\phi(X_{i})\phi^{\rho}(Y_{i}))$, and letting $\omega(S)=\prod_{i\in S}\tfrac{\sigma_{i}}{\sigma_{i}^{\rho}}$, we obtain for a given $S\subseteq V$
\begin{align*}
\E_{\pi,\rho}(\phi_{S}(X)\phi^{\rho}_{S}(Y)) =\prod_{i\in S}\text{cor}_{\pi,\rho}(X_{i},Y_{i})=\prod_{i\in S}\frac{\sigma_{i}}{\sigma_{i,\rho}}\rho=\omega(S)\rho^{|S|}
\end{align*}
Using the Fourier representation we obtain 
\begin{align*}
\E_{\pi,\rho}(f(X)f(Y))= \sum_{S\subseteq V}  \omega(S)\rho^{|S|}\hat{f}^{\pi}(S)\hat{f}^{\pi,\rho}(S)
\end{align*}
The mean of $f(X)$ is $\hat{f}^{\pi}(\varnothing)$ and for $f(Y)$ the mean is $\hat{f}^{\pi,\rho}(\varnothing)$, and so we obtain the covariance.
\end{proof}

%---------------------------------------------------------------------------
\subsection{Approximate decision reliability}\label{proof-sec:approx-dec}
\noindent
{\bf Proposition \ref{prop:approx-decision-reliability}} 
Let $f$ be an LTF Boolean function and $Y$ be an $n$ vector defined in (\ref{eq:measurement-error}) such that the correlation is $\text{cor}_{p_{i}\rho}(X_{i},Y_{i})=\omega(i)\rho$ for all $i$. Then the decision reliability is the square of
\begin{align*}
\text{cor}_{\pi,\rho}(f(X),f(Y)) = \rho\sum_{i\in V} \omega(i)\hat{f}^{\pi}(i)\hat{f}^{\pi,\rho}(i)+ O(\rho^{2}/2)
\end{align*}
\begin{proof}
From Lemma \ref{lem:variance-covariance-f(x)-f(y)} we obtain that
\begin{align}\label{eq:decision-reliability}
\text{cor}_{\pi,\rho}(f(X),f(Y))= \frac{ \sum_{S\ne \varnothing} \omega(S)\rho^{|S|}\hat{f}^{\pi}(S)\hat{f}^{\pi,\rho}(S)}
	{\sqrt{1-\hat{f}^{\pi}(\varnothing)}\sqrt{1-\hat{f}^{\pi,\rho}(\varnothing)}}
\end{align}
The proof follows immediately from noting that since $0\le \rho\le 1$, so that the terms in the covariance for $|S| < 2$ will be small. And since all terms in the Fourier coefficients are bounded, we obtain the bound for $\rho^2$ as the largest terms. Because $f$ is an LTF, we additionally have that the Fourier coefficients $\hat{f}^{\pi}(S)$ and $\hat{f}^{\pi,\rho}(S)$ will be at most $1/2$ \citep[][Theorem 5.2]{ODonnell:2014}. Hence, the bound is accurate.
\end{proof}
%

%%
%\begin{align*}
%\E_{p_{i},\rho}(\phi^\rho(X_{i})\phi^{\rho}(Y_{i})) 
%&= \underbrace{\frac{1}{4}(1+\rho)^2\E_{p_{i}}(\phi^\rho(X_{i})\phi^{\rho}(Y_{i}))}_{\text{both not flipped}} 
%+\underbrace{ \frac{1}{4}(1-\rho)^2\E_{p_{i}}(\phi^\rho(-X_{i})\phi^{\rho}(-Y_{i}))}_{\text{both flipped}}\\
%&+\underbrace{\frac{1}{4}(1-\rho)(1+\rho)\E_{p_{i}}(\phi^\rho(X_{i})\phi^{\rho}(-Y_{i}))}_{\text{$Y$ flipped}}
%+\underbrace{\frac{1}{4}(1-\rho)(1+\rho)\E_{p_{i}}(\phi^\rho(-X_{i})\phi^{\rho}(Y_{i}))}_{\text{$X$ flipped}}\\
%&= \underbrace{\frac{1}{4}(1+\rho)^2\E_{p_{i}}(\phi^\rho(X_{i})\phi^{\rho}(Y_{i}))}_{\text{both not flipped}} +\underbrace{ \frac{1}{4}(1-\rho)^2\E_{p_{i}}(\phi^\rho(X_{i})\phi^{\rho}(-Y_{i}))}_{\text{both flipped}}\\
%\end{align*}
%%
%And 
%%
%\begin{align*}
%\E_{p_{i}}(\phi^\rho(X_{i})\phi^{\rho}(Y_{i})) 
%=
%\text{cor}_{p_i}(X_i,Y_i)+(1-\rho)^2\frac{\mu_{i,X}\mu_{i,Y}}{\sigma_{i,\rho,X}\sigma_{i,\rho,Y}}
%=: \psi_{XY}
%\end{align*}
%%
%And so
%%
%\begin{align*}
%\E_{p_{i},\rho}(\phi^\rho(X_{i})\phi^{\rho}(Y_{i})) 
%&= \underbrace{\frac{1}{4}(1+\rho)^2\psi_{XY}}_{\text{both not flipped}} 
%+\underbrace{ \frac{1}{4}(1-\rho)^2\psi_{XY}}_{\text{both flipped}}
%-\underbrace{\frac{1}{4}(1-\rho)(1+\rho)\psi_{XY}}_{\text{$Y$ flipped}}
%-\underbrace{\frac{1}{4}(1-\rho)(1+\rho)\psi_{XY}}_{\text{$X$ flipped}}\\
%&=\frac{2}{4}(1-\rho)(1+\rho)\psi_{XY}
%\end{align*}
%%

%---------------------------------------------------------------------------
\section{Proofs of supporting lemmas}\label{sec:proofs-support}\noindent

%---------------------------------------------------------------------------
\subsection{May's theorem}\label{proof-sec:may-thm}
\noindent
{\bf Theorem \ref{lem:majority-properties}} \citep{May:1952} 
The majority function $\text{maj}_{n}:\{-1,1\}^{n}\to \{-1,1\}$ defined by $\text{maj}_{n}(x)=\text{sgn} \sum_{\i=1}^{n} x_{i}$ has the properties (a)-(e) below. A function $f$ is 
\begin{itemize}
\item[(a)] {\em monotone} or is {\em positively responsive} if for $x\le y$ $(x_{j}\le y_{j} \forall j)$ implies that $f(x)\le f(y)$, and 
\item[(b)] {\em odd} or {\em neutral} if $f(-x)=-f(x)$;
\item[(c)] {\em unanimous} if $f(-1,-1,\ldots,-1)=-1$ and $f(1,1,\ldots,1)=1$;
\item[(d)] {\em symmetric} or {\em anonymous} if for any permutation $\pi:\{-1,1\}^{n}\to \{-1,1\}^{n}$ of the coordinates in $x$ we have $f(x^{\pi})=f(x)$;
\item[(e)] {\em transitive-symmetric} if for any $i\in V$ there is a permutation $\pi:\{-1,1\}^{n}\to \{-1,1\}^{n}$ of the coordinates in $x$ that puts $x_{i}$ in place of $x_{j}$, such that $f(x^{\pi})=f(x)$.
\end{itemize}
\begin{proof}{}
(a) The majority function is monotone because if $x_{i}\le y_{i}$ for all $i$, then $\sum_{i}x_{i}\le \sum_{i} y_{i}$, and by consequence $\text{sgn}\sum_{i}x_{i}\le \text{sgn}\sum_{i}y_{i}$. (b) Since for $x_{i}$ in the $\{-1,1\}$ domain multiplying by $-1$ is the negation of $x_{i}$, we have that $\text{sgn}(-x_{1}-x_{1}-\cdots-x_{n})=-\text{sgn}(x_{1}+x_{2}+\cdots+x_{n})$, where we obtain the negation of the sign function. (c) Follows from (a). (d) Because the majority function only considers the sum, any permutation $x^{\pi}$ of $x$ will have $\sum_{i}x_{i}=\sum_{i}x^{\pi}$. Finally, (e) is implied by (d) because (e) refers to particular permutations while (d) is about any permutation.
\end{proof}
%

%---------------------------------------------------------------------------
\subsection{Conditional product measure}\label{proof-sec:conditional-product-measure}
\noindent
{\bf Lemma \ref{lem:conditional-product-measure}}
Let $\Prob(x)$ be the joint distribution of the Ising model with probability of $x$ in $\{-1,1\}^{n}$ as in (\ref{eq:ising-joint-probability}). Furthermore, in the conditional probability $\Prob_{i}$ we use the parameterisation $\theta_{ij}/2$, and let $Z_{\pi}(x)=\prod_{i\in V} Z_{i}(x_{\partial i})$. Then we have the factorisation of graph $G$ 
\begin{align}\label{eq:probability-joint-conditional}
\Prob(x) =  \frac{Z_{1}(x_{\partial 1})}{Z_{V}^{\frac{1}{n}}}\Prob(x_{1}\mid x_{\partial 1})\times \cdots\times  \frac{Z_{n}(x_{\partial n})}{Z_{V}^{\frac{1}{n}}}\Prob(x_{n}\mid x_{\partial n})
\end{align}

\begin{proof}
The statement about the probability for each configuration $x$ is easy to see. The product 
\begin{align*}
\Prob(x) =   \frac{Z_{1}(x_{\partial 1})}{Z_{V}^{\frac{1}{n}}}\Prob(x_{1}\mid x_{\partial 1})\times \cdots\times  \frac{Z_{n}(x_{\partial n})}{Z_{V}^{\frac{1}{n}}}\Prob(x_{n}\mid x_{\partial n})\end{align*}
gives the normalising constant $Z_{V}$ and the product of conditionals with the normalising term removed by $Z_{p_{i}}(x_{\partial i})$, leads to
\begin{align*}
\prod_{i\in V} \exp\left( \xi_{i}x_{i} + \frac{1}{2}x_{i}\sum_{j\in \partial i}x_{j}\right)=
\exp\left( \sum_{i}\xi_{i}x_{i} +2\frac{1}{2}\sum_{(i,j)\in E} x_{i}x_{j}\right)
\end{align*}
because each node is visited twice and all neighbourhoods $\partial 1, \partial 2,\ldots , \partial n$ together give the edge set $E$. 
\end{proof}
%

%---------------------------------------------------------------------------
\subsection{Influence for $p$-biased and dependent variables}\label{proof-sec:p-biased-influence}
\noindent
{\bf Lemma \ref{lem:p-biased-influence}}
For a Boolean function $f:\{-1,1\}^{n}\to \{-1,1\}$ with the variables $x_{i}$ having mean $\mu_{i}=2p_{i}-1$ and variance $\sigma^{2}_{i}=1-\mu_{i}^{2}$, the influence for item $i$ is
\begin{align*}
\mathbb{I}_{i}^{\pi}(f) = \sum_{S\ni i}\hat{f}^{\pi}(S)^{2}
\end{align*}
where $S\ni i$ denotes the subsets $S\subseteq V$ such that $i\in S$. Furthermore, for monotone Boolean functions $f$ $\mathbb{I}_{i}^{p_{i}}(f)=\sigma_{i}\mathbb{I}_{i}(f)$.
\begin{proof}
Following the definition of influence, we obtain
\begin{align*}
\mathbb{I}_{i}^{\pi}(f) = \E_{\pi} (D_{\phi,i} f(X)^{2}) = \E_{\pi} \left(\sum_{S\ni i}\hat{f}^{\pi}(S)\phi_{S\backslash \{i\}}(X)\sum_{S \ni i}\hat{f}^{\pi}(T)\phi_{T\backslash \{i\}}(X) \right)
\end{align*}
\begin{align*}
\E_{\pi} \phi_{S\backslash \{i\}}(X)\phi_{T\backslash \{i\}}(X) =
\begin{cases}
1		&\text{ if } S=T\\
0		&\text{ if } S\ne T
\end{cases}
\end{align*}
because $\E_{p_{i}}\phi(X_{i})^{2}=1$ if $i=j$, and 0 if $i\ne j$. Hence,
\begin{align*}
\mathbb{I}_{i}^{\pi}(f) = \sum_{S\ni i}\hat{f}^{\pi}(S)^{2}
\end{align*}
Furthermore, in terms of the unbiased version, $D_{\phi,i}f=\sigma_{i}D_{i}f$. The result now follows for monotone functions $f$. 
\end{proof}
%

%---------------------------------------------------------------------------
\subsection{Influence of monotone functions}\label{proof-sec:influence-fourier-s1}
\noindent
{\bf Lemma 6}
%\ref{lem:influence-fourier-s1}} 
The influence $\mathbb{I}_{i}^{\pi}(f)$ of $i\in V$ on function monotone $f:\{-1,1\}^{n}\to \{-1,1\}$ can be described in terms of the Fourier coefficients $\hat{f}^{\pi}(S)$ by singleton sets $S=\{i\}$
\begin{align*}
\mathbb{I}_{i}^{\pi}(f) = \frac{1}{\sigma_{i}}\hat{f}^{\pi}(i)
\end{align*}
where $\hat{f}^{\pi}(i)=\hat{f}^{\pi}(\{i\})$.

\begin{proof}
This Lemma is an extension of \citet[][Proposition 2.21]{ODonnell:2014}, where we include that the $X_i$ are $p$-biased and dependent. 

Because $f$ is monotone by assumption, we have that $D_{i}f\ge 0$, and so $D_{i} f(X) =D_{i} f(X)^{2}$ and is either 0 or 1, hence $\mathbb{I}_{i}(f) = \E_{\pi} (D_{i} f(X))$. And
\begin{align*}
\E_{\pi} (D_{i} f(X)) = \E_{\pi} \left(  \sum_{\underset{i\in S}{S\subseteq V}} \hat{f}^{\pi}(S)\phi_{S\backslash \{i\}}(X) \right) =  \sum_{i\in S}\hat{f}^{\pi}(S)\E_{\pi}(\phi_{S\backslash \{i\}}(X))
\end{align*}
And 
\begin{align*}
\E_{\pi}(\phi_{S\backslash \{i\}}(X))=
\begin{cases}
1	&\text{ if } i\in S=\{i\}\\
0	&\text{ if } i\notin S
\end{cases}
\end{align*}
because if $i\in S$ then $S\backslash \{i\}=\varnothing$ only if $S=\{i\}$. Hence, we have nonzero coefficients with $\hat{f}^{\pi}(S)=\hat{f}^{\pi}(\{i\})$.
\end{proof}
%

%---------------------------------------------------------------------------
%\begin{lemma}\citep[][Proposition 2.32]{ODonnell:2014}\label{lem:expectation-agree}
%Let $f:\{-1,1\}^{n}\to \{-1,1\}$ be any Boolean function for unbiased variables $X_{i}$ and $W$ is the number of items that agree with $f$. Then
%%
%\begin{align*}
%\E(W) = \frac{n}{2}+\frac{1}{2}\sum_{i=1}^{n}\hat{f}(i)
%\end{align*}
%%
%\end{lemma}
%%
%\begin{proof}
%Because
%%
%\begin{align*}
%\sum_{i=1}^{n}\hat{f}(i) = \sum_{i=1}^{n}\E(f(X)X_{i}) = \E(f(X)(X_{1}+\cdots + X_{n}))
%\end{align*}
%%
%and $f(x)(x_{1}+\cdots + x_{n})$ is the the number of items that agree with $f(x)$ or with $-f(x)$, and so is $w-(n-w)=2w-n$. Hence, 
%%
%\begin{align*}
%\sum_{i=1}^{n}\hat{f}(i) = \sum_{i=1}^{n}\E(f(X)X_{i}) = \E(f(X)(X_{1}+\cdots + X_{n}))=\E(2W-n)=2\E(W) - n
%\end{align*}
%%
%leading to the result.
%\end{proof}
%%

%---------------------------------------------------------------------------
\section{Estimation of graph and Fourier coefficients}\label{sec:algorithm-fourier-analysis}\noindent
The $z$-transformation $\phi(x_{i})=(x_{i}-\mu_{i})/\sigma_{i}$ for all $i\in V$ requires the mean (\ref{eq:p-biased-mean}) and variance (\ref{eq:p-biased-variance}), which are determined by the probabilities $p_{i}=\Prob(X_{i}=1)$. Because we assume a graph for the items $X_{i}$ we can model the probability $p_{i}$ by the nodes in the neighbourhood $\partial i$, the nodes that are directly connected to node $i$. For each node $i\in V$ we estimate the probability $p_{i}$ that $X_{i}=1$ using the Ising model by
\begin{align}
\hat{p}_{i}=\frac{\exp(\hat{\xi}_{i}+\sum_{j\in \partial i}\hat{\theta}_{ij}x_{ij})}{1+\exp(\hat{\xi}_{i}+\sum_{j\in \partial i}\hat{\theta}_{ij}x_{ij})}
\end{align}
where $\hat{\xi}_{i}$ and $\hat{\theta}_{ij}$ are estimates of the $\xi_{i}$ and $\theta_{ij}$, respectively. Note that we only use local information with respect to the graph in that the nodes $j$ in the neighbourhood $\partial i$ are responsible for determining the probability of node $i$. Obtaining the estimate $\hat{p}_{i}$ for each node $i\in V$ we can determine $\hat{\mu}_{i}=2\hat{p}_{i}-1$ and $\hat{\sigma}_{i}=1-\hat{\mu}_{i}^{2}$. With these estimates we obtain 
\begin{align}
\hat{\phi}(x_{i})= \frac{x_{i} - \hat{\mu}_{i}}{\hat{\sigma}_{i}}
\end{align}
which we can plug in the $p$-biased Fourier coefficient $\hat{f}^{\hat{\pi}}(S)$ in (\ref{eq:p-biased-coefficient}) with $\hat{\pi}=(\hat{p}_{1},\ldots, \hat{p}_{n})$. 

To obtain the estimates $\hat{\xi}_{i}$ and $\hat{\theta}_{ij}$ for all $i$ and $j$ in $V$ we use the lasso penalty on the conditional distributions \citep{Geer:2008,Ravikumar:2010,Borkulo:2014}. The lasso version for logistic regression optimises the pseudo-likelihood \citep{Besag:1974}. Choose node $i\in V$ and let the logit function for $p_{i}=\Prob(X_{i}=1)$ be 
\begin{align}\label{eq:lasso-optimisation}
g_{i}(x_{t}) = \log \left( \frac{p_{i}}{1-p_{i}} \right) = \xi_{i} +\sum_{j\in \partial i}\theta_{ij}x_{ij,t}
\end{align}
for observation $x_{t}$ with $t\in U$ the observation units. 
Then the pseudo-likelihood is then 
\begin{align}
\min_{\beta \in\mathbb{R}^{n}}\frac{1}{|U|}\sum_{t\in U}\Bigl(-x_{i}g_{i}(x_{t}) + \log(1 + \exp(g_{i}(x_{t}))\Bigr) +\lambda ||\beta_{i} ||_{1}
\end{align}
where $\lambda >0$ is the penalty parameter and $\beta_{i}=(\xi_{i},\theta_{ij};j\in V\backslash \{i\})$ is the parameter vector for node $i$ and $||\beta_{i} ||_{1}=\sum_{j=1}^{n}|\beta_{i,j}|$ is the $\ell_{1}$ norm. This function is convex and so can be optimised using for instance the coordinate descent algorithm, where each parameter $\beta_{i,j}$ (coordinate) is optimised in turn \citep{Hastie:2015,Waldorp:2019}. We estimate the parameters $\beta_{i}$ for each node $i$ in turn. 

Once we have estimates $\hat{\beta}_{i}$ for all nodes $i\in V$ we have estimates $\hat{p}_{i}$ of the probabilities for all nodes and so the transformation $\phi(x_{i})$, all based on the conditional distributions. We can then compute for each subset $S\subseteq V$ the functions 
\begin{align}
\hat{\phi}_{S}=\prod_{i\in S}\hat{\phi}(x_{i})
\end{align}
where we plugged in the estimates $\hat{\phi}$ in (\ref{eq:basis-function-orthogonal-p-biased}). This leads to the estimates of the Fourier coefficients
\begin{align}\label{eq:fourier-coefficient-sample-estimate}
\hat{f}_{U}^{\hat{\pi}}(S)=\frac{1}{|U|}\sum_{t\in U}f(x_{t})\hat{\chi}^{\phi}_{S}(x_{t})
\end{align}
Note that by Proposition \ref{prop:approx-decision-reliability} we require only the first few coefficients, singleton sets $S=\{i\}$ and duo sets $S=\{i,j\}$ because the effect of the higher order sets is small.

The Fourier coefficients are required to compute the stability and noise sensitivity. The definition of stability in (\ref{eq:stability}) uses both the original values in $x$ and the ones with measurement error $y$; and we of course only have those with measurement error.

\vspace{1em}
\noindent
{\bf Statistical guarantees}\label{sec:statistical-guarantees}\\\noindent
Here we investigate the rate of convergence of the estimates of the Fourier coefficients we can expect based on estimation of the Ising parameters. 

Our algorithm in above involves estimation of the Ising parameters $\beta_{i}=(\xi_{i},\theta_{ij},j\in V\backslash \{i\})$ with the lasso. The lasso is known to have a difficult distribution \citep{Potscher:2009c} and hence we cannot directly use this to obtain bounds. In order to obtain convergence rates on the estimate, we use results from the so-called desparsified lasso \citep{Geer2013,Javanmard:2014}, where a projection of the residuals is added to `desparsify' the lasso (make the 0s non-zero again based on the residuals). 

The lasso has a soft threshold such that parameter values within the range of the penalty $\lambda$ of 0 will be set to exactly 0. This implies that the sampling distribution of the lasso estimate has unit mass at these points, which destroys the nice property of the sampling distribution \citep{Potscher:2009c}, usually obtained with the central limit theorem. The desparsified lasso projects the residuals from the lasso based on an approximation of the inverse of the second order derivative of the optimisation function in (\ref{eq:lasso-optimisation}). For each $\beta_{i}$ in the list of nodewise optimisations, let $X_{t}$ denote the $t$th observation of the $n-1$ items (without item $i$), and let $p_{i}(X_{t})$ denote the conditional probability in (\ref{eq:ising-conditional-probability}) with the $t$th observation for the $n-1$ remaining items plugged in. Furthermore, let the function for the nodewise regression with respect to item $i$, without the lasso penalty, be
\begin{align}
\psi(\beta_{i}) = -x_{it}m_{i}(x_{t}) + \log(1+\exp(m_{i}(x_{t})))
\end{align}
where $x_{it}$ is the value from observation $t$ of item $i$. Then we have the $n\times n$ second order derivative matrix for the $n$ parameters in $\beta=(\xi_{i},\theta_{ik},k\in V\backslash\{i\})$
\begin{align}\label{eq:second-deriv}
\nabla^{2}\psi(\beta_{i})=\frac{1}{m}\sum_{j=1}^{m}\E_{\pi}( p_{i}(X_{j})p_{i}(-X_{j})X_{j}X_{j}^{\sf T})
\end{align}
We denote this theoretical second order derivative by $\Sigma_{i}=\nabla^{2}\psi(\beta_{i})$ and assume that its eigenvalues are $>0$ \citep{Geer2013,Waldorp:2019}. We obtain an estimate $\hat{\Sigma}_{i}$ by removing the expectation operator in (\ref{eq:second-deriv}). We often have that $\hat{\Sigma}$ is singular, certainly so when $n>m$. Therefore, in general we construct an approximate inverse $\hat{\Theta}_{i}$ such that the difference $||\hat{\Sigma}_{i}\hat{\Theta}_{i}-I_{n}||_{\infty}=O_{p}(\sqrt{\log(n)/m})$, where $||\cdot ||_{\infty}$ is the max norm and $O_{p}(v)$ means that the random variable $V_{m}$ is for all $\varepsilon$, $\Prob( |V_{m}| \le K_{\varepsilon} b_{m})>1-\varepsilon$ for $m\to\infty$ and some $K_{\varepsilon} >0$, i.e., $V_{m}/b_{m}$ is bounded in probability by $K_{\varepsilon}$ \citep[see, e.g., ][]{Geer2013,Javanmard:2014}. Then we can construct the desparsified lasso by 
\begin{align}\label{eq:desparsified-lasso}
\hat{\beta}_{i}^{dL} = \hat{\beta}_{i} + \hat{\Theta}_{i}\nabla \psi(\beta_{i})
\end{align}
where $\nabla\psi(\beta_{i})$ is the $n$ vector of first order derivatives of $\psi$ with respect to $\beta_{i}$ with $j$th element
\begin{align}
\nabla_{j} \psi(\beta_{i}) = \frac{1}{m}\sum_{t=1}^{m}(-x_{jt}+p_{i}(x_{t}))x_{jt}
\end{align}
The matrix $\hat{\Theta}_{i}$ can be obtained, for instance, by performing nodewise regressions on the remaining (i.e., predictor) items \citep[see, e.g., ][]{Geer:2014}. Then we obtain the approximation for the desparsified lasso   
\begin{align}\label{eq:asymptotic-representation-desaprsified-lasso}
\hat{\beta}_{i}^{dL} = \beta_{i} + \frac{1}{\sqrt{m}} Z_{i} + o_{p}(1)
\end{align}
where the $n$ vector $Z_{i}$ is a normal random variable with mean 0 and variance matrix
$\text{var}(Z_{i}) = \hat{\Theta}_{i}\Sigma_{i}\hat{\Theta}_{i}$ and $o_{p}(1)$ means a random variable that converges in probability to 0, i.e., for every $\varepsilon>0$, $\Prob(|V_{m}|\le \varepsilon)> 1-\varepsilon$ as $m\to\infty$. We assume here that the lasso estimate $\hat{\beta}_{i}$ is obtained with penalty $\lambda\ge \sqrt{\log(p)/n}$ \citep{Geer:2014}.

With the representation in (\ref{eq:asymptotic-representation-desaprsified-lasso}) we can obtain bounds on the estimation error of the Fourier coefficients. We first plug in the representation (\ref{eq:asymptotic-representation-desaprsified-lasso}) in the conditional distribution $p_{i}$ in (\ref{eq:ising-conditional-probability}). For the conditional probabilities we can plug in the representation and obtain 
\begin{align*}
p_{i}(\hat{\beta}_{i}^{dL}) = \text{logit}(\tilde{x}_{t}^{\sf T}\beta_{i}+Z_{i}/\sqrt{m}+o_{p}(1))
\end{align*}
where $\tilde{x}_{t}$ is the $t$th observation $(1,x_{t})$, where the 1 is included for the threshold parameter $\xi_{i}$, and the $\text{logit}(z)$ function is $1/(1+\exp(-z))$. We then see that we have in the denominator 
\begin{align*}
\exp(-\tilde{x}^{\sf T}\beta_{i})\exp(Z_{i}/\sqrt{m}+o_{p}(1))
\end{align*}
The second term $\exp(Z_{i}/\sqrt{m}+o_{p}(1))$ converges with $m$ to 1 at the rate of $1/\sqrt{m}$. Hence, the conditional probability $p_{i}(\hat{\beta}_{i}^{dL})$ converges to the true conditional probability $p_{i}$ at rate $1/\sqrt{m}$. The  transformation $\hat{\phi}$ using $\hat{\beta}_{i}^{dL}$ with the mean $\mu_{i}=2p_{i}-1$ and standard deviation $\sigma_{i}=\sqrt{1-\mu_{i}^{2}}$, do not alter the convergence in probability, and hence, we obtain $\hat{\phi}(x_{i})=\phi(x_{i})+o_{p}(1)$. And finally, the computation of the Fourier coefficient (of order 1) in (\ref{eq:fourier-coefficient-sample-estimate}) gives 
\begin{align*}
\hat{f}^{\pi}_{m}(i) = \frac{1}{m}\sum_{t=1}^{m}f(X_{t})\hat{\phi}(X_{it})=\frac{1}{m}\sum_{t=1}^{m}f(X_{t})\phi(X_{it}) + o_{p}(1)
\end{align*}
Because the convergence of the average of the Fourier coefficient with the correct transformation with rate $1/\sqrt{m}$, we find that the Fourier coefficient converges at the reasonable rate $1/\sqrt{m}$. This convergence rate is for the coefficients of order 1, for higher order coefficients with $\phi_{S}$ where $|S|=k$, say, we obtain slower convergence due to the product in $\phi_{S}$.

%---------------------------------------------------------------------------
\end{document}